\documentclass[twocolumn,10pt]{article}

\usepackage[margin=0.75in]{geometry}
\usepackage{times,newtxmath,mathrsfs} 
\usepackage[T1]{fontenc}
\usepackage[utf8]{inputenc}
\usepackage[small]{titlesec} 
\usepackage{balance}

\usepackage{abstract} 

\usepackage{color}
\usepackage{graphicx}
\usepackage[font={sf,bf,small}]{caption}
\usepackage{booktabs}
\usepackage{supertabular} 
\usepackage{amssymb,amsmath,bm} 

\usepackage{smartdiagram}
\usepackage{tikz}
\usetikzlibrary{shapes.geometric, arrows}
\tikzstyle{end} = [ellipse, minimum width=3cm, minimum height=0cm, text centered, draw=black, fill=none]
\tikzstyle{process2} = [rectangle, minimum width=3cm, minimum height=0cm, text centered, draw=black, fill=none]
\tikzstyle{decision} = [diamond, minimum width=0cm, minimum height=0cm, text centered, draw=black, fill=none, aspect=2]
\tikzstyle{arrow} = [thick,->,>=stealth]

\makeatletter
\newcommand{\vast}{\bBigg@{4}}
\newcommand{\Vast}{\bBigg@{5}}
\makeatother

\usepackage{fancyhdr}
\usepackage{lastpage}
\usepackage[numbers,square,sort,comma]{natbib}
\usepackage{abstract}

\usepackage[colorlinks=true,allcolors=blue,breaklinks=true]{hyperref} 

\def\doubleunderline#1{\underline{\underline{#1}}}

\title{\textbf{\textsf{Hydrodynamic Bulge Testing: Materials Characterization without Measuring Deformation}}}

\author{\textbf{\textsf{Vishal Anand}}\\
        \textsf{\small School of Mechanical Engineering,}\\
        \textsf{\small Purdue University,}\\
        \textsf{\small West Lafayette, Indiana, 47907, USA}\\[2mm]
        \textbf{\textsf{Sanjan C.\ Muchandimath}}\\
        \textsf{\small Department of Aerospace Engineering,}\\
        \textsf{\small Indian Institute of Technology Madras,}\\
        \textsf{\small Chennai 600036, India}\\[2mm]
        \textbf{\textsf{Ivan C.\ Christov}}\thanks{Author to whom correspondence should be addressed.}\\
        \textsf{\small School of Mechanical Engineering,}\\
        \textsf{\small Purdue University,}\\
        \textsf{\small West Lafayette, Indiana, 47907, USA}\\
        \textsf{\small e-mail: christov@purdue.edu}
}

\date{}

\begin{document}

\twocolumn[
  \begin{@twocolumnfalse}
    \maketitle
    \vspace{-1cm}
    \begin{abstract}
      \noindent
      \textit{Characterizing the elastic properties of soft materials through bulge testing relies on accurate measurement of deformation, which is experimentally challenging. To avoid measuring deformation, we propose a hydrodynamic bulge test for characterizing the material properties of thick, pre-stressed elastic sheets via their fluid--structure interaction with a steady viscous fluid flow. Specifically, the hydrodynamic bulge test relies on a pressure drop measurement across a rectangular microchannel with a deformable top wall. We develop a mathematical model using first-order shear-deformation theory of plates with stretching, and the lubrication approximation for Newtonian fluid flow. Specifically, a relationship is derived between the imposed flow rate and the total pressure drop. Then, this relationship is inverted numerically to yield estimates of the Young's modulus (given the Poisson ratio), if the pressure drop is measured (given the steady flow rate). Direct numerical simulations of two-way-coupled fluid--structure interaction are carried out in ANSYS to determine the cross-sectional membrane deformation and the hydrodynamic pressure distribution. Taking the simulations as ``ground truth,'' a hydrodynamic bulge test is performed using the simulation data to ascertain the accuracy and validity of the proposed methodology for estimating material properties. An error propagation analysis is performed via Monte Carlo simulation to characterize the susceptibility of the hydrodynamic bulge test estimates to noise. We find that, while a hydrodynamic bulge test is less accurate in characterizing material properties, it is less susceptible to noise, in the input (measured) variable, than a hydrostatic bulge test.\\[2mm]
      }
    \end{abstract}
    \bigskip
    \bigskip
  \end{@twocolumnfalse}
]
\saythanks 

\section{Introduction}
\label{sec:Intro}
Bulge testing is a standard technique for measuring mechanical properties of thin films of elastic materials \cite{VN92,SN92}. In the development of microfluidic platforms, soft polymeric materials, such as polydimethylsiloxane (PDMS) \cite{MW02}, are used for rapid manufacture of fluid-conveying microchannels \cite{ALA99,ACJCDWWW00} via soft lithography \cite{XW98}. However, the mechanical properties (such as the Young's modulus and Poisson ratio) of such materials are sensitive to how the polymers are mixed, how long the mixture is cured, and the ambient thermal conditions \cite{JMTT14}. Therefore, bulge testing is used to estimate the elastic properties of soft materials, such as PDMS and also polyurethane (PU), used in microfluidics \cite{J08,Hetal18}. 

A bulge test involves clamping a thin elastic sheet over an orifice (or a window) and, then, measuring its deformation under a known (usually uniform) pressure field \cite{SN92,J08,Hetal18}. The measured deformation as a function of the known pressure load can then be converted to strain as a function of stress, by employing a suitable structural mechanics model (e.g., the theory of linear elasticity). In turn, knowing the stress as a function of the imposed strain allows for straightforward estimation of the elastic modulus of the material (assuming that the Poisson's ratio is known) \cite{SN92,J08,Hetal18}. Knowledge of  the stress distribution within the structure is also used to estimate fracture properties of the material \cite{ZYSLYL08,ZYLJYY09,YP02}. Several techniques have been proposed to improve the accuracy of ``traditional'' bulge tests. These improvements include, but are not limited to, accounting for the film's bending stiffness \cite{NHG12}, accounting for pre-stress in the film \cite{SN92,VN92}, considering the possibility of buckling \cite{SVOH18} and better prediction of the stress distribution near edges \cite{YLMW14} by using elastically clamped (instead of the traditional rigidly clamped) boundary conditions \cite{ZYSLYL08,ZYLJYY09}. 

One of the main sources of uncertainty in bulge tests is the experimental measurement of the film's deformation \cite{Hetal18}. Traditionally,  deformation has been measured by interferometric techniques and, less frequently, by high-resolution microscopy. Both of these measurement techniques have certain limitations.  On the one hand, interferometers are prone to errors induced from external sources of vibrations \cite{Hetal18}, which limits the spatial resolution of the measurements and makes it difficult to accurately resolve deformations in the small-strain regime relevant to bulge testing \cite{SN92,Hetal18}. On the other hand, microscopes are not well suited to analyze samples with high reflectance, such as PDMS \cite{Hetal18}. Thus, there is motivation for developing bulge testing techniques that bypass the deformation measurement altogether.

Often, bulge testing techniques discussed in the literature have focused on circular membranes \cite{SN92}, with only a few studies addressing the case of rectangular membranes with pre-stress \cite{ZYSLYL08, ZYLJYY09,YP02,ZPMSB98} using energy minimization methods \cite{VN92}. Residual pre-stress (pre-tension) is common in  samples being tested because the thin film of material has to be stretched taut over an orifice (say, a rectangular microchannel) to ensure that it is flat before the commencement of the experiment \cite{DHTJ17,BGB18}. Furthermore, most of the bulge testing theories in the literature assume that the film has negligible thickness and deformations due to shear along the transverse direction are, thus, not accounted for. 

To improve upon some of these drawbacks of static bulge testing, we propose a theory of \emph{hydrodynamic} bulge tests, in which the applied pressure load on the thin structure is due to viscous fluid flow underneath it. We account for both uniform isotropic pre-stress and the finite thickness of a rectangular elastic sheet. The novelty of this approach is that it does not require a measurement of the deformation profile of the elastic membrane. Through this approach, we are able to characterize the elastic properties of a soft material using a mathematical model derived to relate the \emph{total} pressure drop, at steady state, over the length of the elastic sheet to the imposed volumetric flow rate of the fluid flow underneath it.

The interplay of pre-stress-induced stretching, pressure-induced bending and the finite thickness of a plate-like structure leads to several different physical regimes of flow-induced deformation. Thus, a hydrodynamic bulge test is an example of low-Reynolds-number \textit{fluid--structure interaction} (FSI) \cite{DS16}. This problem, rather than the problem of the deflection of circular membranes typically studied in the bulge testing literature, is more relevant to microfluidics because PDMS microchannels' walls are generally not circular but rectangular \cite{GEGJ06,Hetal18}. A mathematical model of such FSI requires the use of the lubrication approximation to obtain the leading-order (with the flow-wise aspect ratio as the small parameter) fluid flow field, and then coupling it to a deformation profile obtained under an appropriate structural mechanics model (herein, a plate theory) \cite{GEGJ06,CCSS17, MY19}. The main result of the mathematical derivations in this work is the flow rate--pressure drop relationship for flow in a long and shallow rectangular microchannel with deformable top wall. In \cite{OYE13}, this relationship was employed to non-invasively measure the non-uniform hydrodynamic pressure distribution (within a microchannel) from the wall deformation. Here, we pose the opposite FSI problem: if the pressure profile is known, can a mathematical model be used to infer the deformation? Then, can the \emph{total} pressure drop be used to infer the material properties of the thin solid film that comprises the deformable channel wall? 

To answer these questions in the affirmative, in Sec.~\ref{sec:plates}, we first derive the governing equations of a first-order shear-deformation plate theory, incorporating finite transverse thickness and pre-stress. Specifically, for a long and wide geometry, the problem is reduced to two coupled ordinary differential equations (in the spanwise coordinate) for the rotation of the normal and the vertical displacement (Sec.~\ref{sec:plate_reduction}). Three regimes of deformation are delineated, and a solution for the deformation, given an axially-varying pressure load, is found in each regime (Sec.~\ref{sec:solution_deformation}). Section~\ref{sec:fluid} summarizes the hydrodynamics problem under the lubrication approximation for viscous flow in slender geometries, and we obtain the flow rate--pressure drop relation by coupling the fluid and solid mechanics problems. In Sec.~\ref{sec:computational}, we compare the latter theoretical result to direct numerical simulations of FSI in ANSYS, showing good agreement. On the basis of this validation, a hydrodynamic bulge testing theory is proposed, and a sensitivity (error propagation) analysis is performed on it via Monte Carlo simulations in Sec.~\ref{sec:sensitivity}. Conclusions are stated in Sec.~\ref{sec:conclusion}, and {three} appendices {(Supplemental Material)} provide further mathematical details: {the derivation of the governing partial differential equations of the thick-plate plate theory (Appendix~\ref{app:plate_theory})}, results regarding the deformation profile in different regimes (Appendix~\ref{app:other_regimes}), and the special case of a ``classical'' thin-plate theory (Appendix~\ref{app:thin}).

\section{Structural Mechanics}
\label{sec:plates}

\begin{figure}
  \centering
  \includegraphics[width=\linewidth]{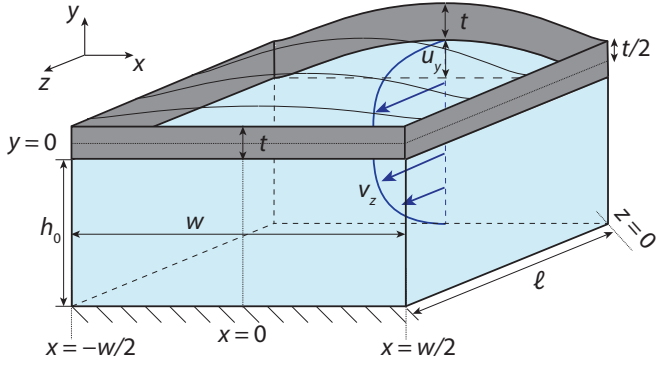}
\caption{Geometry of the problem and notation. A slender elastic membrane (plate) represents the top wall of an otherwise rigid channel. A steady flow is established in the $z$-direction, which gives rise to a pressure gradient that leads to a deformation of the membrane in the $y$-direction. The membrane is clamped on all ends (not shown at $z=0$ for clarity). Reprinted and adapted from Journal of Non-Newtonian Fluid Mechanics 264, Vishal Anand, Joshua David JR, Ivan C.\ Christov, ``Non-Newtonian fluid--structure interactions: Static response of a microchannel due to internal flow of a power-law fluid,'' 62--72 Elsevier \cite{ADC18}, Copyright (2019), with permission from.}
\label{fig:schematic}
\end{figure}

Consider the geometry depicted in Fig.~\ref{fig:schematic}. An elastic plate, clamped on all its edges, is placed as the top wall over a rectangular channel that is long and wide. The plate's thickness is smaller than its spanwise width ($t/w < 1$), but it is not negligible ($t/w\not\to0$). Furthermore, the reference configuration of the plate is assumed to have an uniform (isotropic) pre-tension $T$, defined as a force per unit length (stress resultant). In this section, we {we summarize the key points of a plate theory, based on the Reissner--Mindlin (RM) approach \cite{R45,M51}, that also accounts for the pre-tension/pre-stress/stretching in the elastic body. A more complete discussion is available in Supplemental Material Appendix~\ref{app:plate_theory}.} RM, or ``thick-plate,'' theories are also referred to as first-order shear-deformation theories (FOSDT) (see the recent historical overview of the development of these theories by Challamel and Elishakoff \cite{CE19}). Then, we show how this pre-stressed thick-plate theory can be used to obtain a complete description of hydrodynamic bulge testing of elastic structures.

\subsection{Differential Equations for the Displacement}
To define a FOSDT with both stretching and bending, one obtains substitutes the stress resultants from Eqs.~\eqref{eq:normal_stress_consti}, \eqref{eq:bending_moment_constitutive2} and  \eqref{eq:transverse_shear_res_kappa} from the Supplemental Material into the equations of equilibrium \eqref{eq:equil}, to obtain a set of differential equations for the displacements $(u_{x0},u_y,u_{z0})$ and the rotations of the normal $(\phi_x,\phi_z)$:
\begin{subequations}
\begin{align}
    \label{eq:displacement_in_plane_x}
    D_s\left(\frac{\partial^2 u_{x0}}{\partial x^2}+\nu\frac{\partial^2 u_{z0}}{\partial x\partial z}\right)+\kappa Gt\left(\frac{\partial ^2 u_{x0}}{\partial z^2}+\frac{\partial  ^2u_{z0}}{\partial x\partial z}\right) = 0,\\
\label{eq:displacement_in_plane_z}
     \kappa Gt\left(\frac{\partial^2 u_{z0}}{\partial x^2}+\frac{\partial^2 u_{x0}}{\partial x\partial z}\right)+D_s\left(\frac{\partial ^2 u_{z0}}{\partial z^2}+\nu\frac{\partial  ^2u_{x0}}{\partial x\partial z}\right) = 0,\\
\label{eq:displacement_trans}
     \kappa Gt\left[\left(\frac{\partial ^2 u_{y}}{\partial x^2}+\frac{\partial ^2 u_{y}}{\partial z^2}\right)+\frac{\partial \phi_x}{\partial x}+\frac{\partial \phi_z}{\partial z}\right]+\mathcal{N}+p = 0,\\
\label{eq:rotation_x}
      D_b\left(\frac{\partial^2 \phi_x}{\partial x^2}+\nu\frac{\partial^2\phi_z}{\partial x \partial z}\right)+ D_b\left(\frac{1-\nu}{2}\right)\left(\frac{\partial ^2 \phi_x}{\partial z^2}+\nu\frac{\partial^{2} \phi_z}{\partial x \partial z}\right) \nonumber\\
      -{\kappa G t}\left(\frac{\partial u_{y}}{\partial x}+\phi_x\right) = 0,\\
\label{eq:rotation_z}
     D_b\left(\nu\frac{\partial ^2 \phi_x}{\partial x \partial z}+\frac{\partial ^2\phi_z}{\partial z^2}\right)+D_b\left(\frac{1-\nu}{2}\right)\left(\frac{\partial ^2 \phi_x}{\partial x \partial z}+\frac{\partial^2 \phi_z}{\partial x^2}\right) \nonumber\\
     -\kappa Gt\left(\frac{\partial u_{y}}{\partial z}+\phi_z\right) = 0,
\end{align}\end{subequations}
where $D_b := Et^3/[12(1-\nu^2)]$ is the bending rigidity, and $D_s := Et/(1-\nu)$ is the extensional rigidity, of the plate with Young's modulus $E$ and Poisson ratio $\nu$ \cite{HKO09,reddy04}. {Here, $G := E/[2(1+\nu)]$ is the shear modulus, and  $\kappa$ is Timoshenko's ``shear correction factor'' \cite{T21}, which is  commonly introduced to account for nonuniform distribution of the transverse shear strain across the thickness \cite{GW01,H01,Z06}. Following Zhang \cite{Z06}, and as in previous works \cite{SC18,ADC18}, we take $\kappa = 1$ to ensure consistency of three-dimensional (3D) linear elasticity and RM plate theory in the limit of $t/w \to 0$.}

Equations~\eqref{eq:displacement_in_plane_x} and \eqref{eq:displacement_in_plane_z} completely describe the in-plane displacement field, which is independent of the transverse deflection and/or rotations. In the analysis below, this in-plane displacement field will not be necessary, thus we discard these two equations.   

Finally, in this work, we assume that the stretching response of the plate is due to a known isotropic, uniform pre-tension $T$, i.e., {the normal stress and in-plane shear stress resultants are simply given by}
\begin{equation}
    \label{eq:pre-tension-defined}
    \begin{pmatrix}
    N_{xx} \\ N_{zz} \\  N_{xz}
    \end{pmatrix} = T\begin{pmatrix}
    1 \\ 1 \\ 0
    \end{pmatrix}.
\end{equation}
Then, Eq.~\eqref{eq:displacement_trans} becomes, 
\begin{equation}
\label{eq:thick_tension_dim_final1}
    \kappa G t \left(\frac{\partial \phi_x}{\partial x}+\frac{\partial \phi_{z}}{\partial {z}}\right) + (T+\kappa G t)\left(\frac{\partial ^2 u_{y}}{\partial x^2}+\frac{\partial ^2 u_{y}}{\partial {z^2}}\right) = -p,
\end{equation}
Together, Eqs.~\eqref{eq:thick_tension_dim_final1}, \eqref{eq:rotation_x} and \eqref{eq:rotation_z} describe the transverse deflection and rotations of the normal to the mid-plane of a thick, pre-stressed plate. A  subtle consequence of imposing the pre-stress on the model \textit{a priori}, rather than computing it through internal strains, is that the equations of the present weakly nonlinear theory become equivalent to equations of von K\'arm\'an's plate theory with given constant isotropic pre-tension \cite{HKO09, B10}.

\subsection{Shallow, Slender Plates: Regimes of Deformation}
\label{sec:plate_reduction}

First, we rewrite the governing differential equations \eqref{eq:thick_tension_dim_final1}, \eqref{eq:rotation_x} and \eqref{eq:rotation_z} using the following dimensionless variables:
\begin{multline}
\label{eq:scales}
  X = x/w, \quad Z = z/\ell, \quad  U = u_{y}/\mathcal{U}_c,\\ \Phi_x = \phi_x/\mathcal{F}_x, \quad \Phi_z = \phi_z/\mathcal{F}_z \quad P = p/\mathcal{P}_c,
\end{multline}
Here, $\mathcal{F}_x$, $\mathcal{F}_z$ and $\mathcal{U}_c$ are the characteristic scales for the rotation of the normal in the $x$ and $z$ directions, and the characteristic scale for the deformation itself, respectively. These scales will be determined self-consistently through the mathematical analysis below. The characteristic pressure scale is $\mathcal{P}_c$, which will be obtained from the analysis of the fluid mechanics problem. For a long and wide microchannel, following \cite{CCSS17}, assume that
\begin{equation}
\label{eq:assumption_shallownes}
   h_0 \ll w \ll \ell \qquad \Rightarrow \qquad  \epsilon \ll \delta \ll 1,
\end{equation}
where $\epsilon := {h_0}/{\ell}$ and $\delta := {h_0}/{w}$, and $h_0 $ is the undeformed height of the channel (recall Fig.~\ref{fig:schematic}). Substituting the dimensionless variables from Eq.~\eqref{eq:scales} into Eqs.~\eqref{eq:thick_tension_dim_final1}, \eqref{eq:rotation_x} and \eqref{eq:rotation_z} yields a dimensionless set of governing equations:
\begin{subequations}\begin{multline}
\label{eq:dimless_equilibrium_z_v1}
    \kappa G t\left(\frac{\mathcal{F}_x}{w}\frac{\partial \Phi_x}{\partial X}+\frac{\mathcal{F}_z}{\ell}\frac{\partial \Phi_z}{\partial Z}\right)\\
    +(T+\kappa G t)\left(\frac{\mathcal{U}_c}{w^2}\frac{\partial ^2 U}{\partial X^2 }+\frac{\mathcal{U}_c}{\ell^2}\frac{\partial ^2 U}{\partial Z^2}\right) = - \mathcal{P}_c P,
\end{multline}   
\begin{multline}
\label{eq:dimless_equilibrium_phix_v1}
    \frac{D_b}{w^2}\mathcal{F}_x \frac{\partial^2 \Phi_x}{\partial X^2} + \frac{1}{2}\frac{D_b(1-\nu)}{\ell^2}\mathcal{F}_x\frac{\partial ^2 \Phi_x}{\partial Z^2}\\ +\frac{1}{2}\frac{D_b(1+\nu)}{w\ell}\mathcal{F}_z\frac{\partial ^2 \Phi_z}{\partial X \partial Z}-\kappa G t\left(\mathcal{F}_x\Phi_x + \frac{\mathcal{U}_c}{w}\frac{\partial U}{\partial X}\right) = 0,
\end{multline}
\begin{multline}
\label{eq:dimless_equilibrium_phiz_v1}
    \frac{D_b}{\ell^2}\mathcal{F}_z\frac{\partial^2 \Phi_z}{\partial Z^2} + \frac{1}{2}\frac{D_b(1-\nu)}{w^2}\mathcal{F}_z\frac{\partial ^2 \Phi_z}{\partial X^2} \\
    +\frac{1}{2}\frac{D_b(1+\nu)}{w\ell}\mathcal{F}_x\frac{\partial ^2 \Phi_x}{\partial X \partial Z}-\kappa G t\left(\mathcal{F}_z\Phi_z + \frac{\mathcal{U}_c}{\ell}\frac{\partial U}{\partial Z}\right) = 0.
\end{multline}\end{subequations}
In Eqs.~\eqref{eq:dimless_equilibrium_phix_v1} and \eqref{eq:dimless_equilibrium_phiz_v1}, the terms involving $\partial U/\partial X$ and $\partial U/\partial Z$ arise from the transverse shear resultant, meaning they are a key aspect of the FOSDT. To retain these terms asymptotically, we take
\begin{equation}
\label{eq:phi_scale}
 \mathcal{F}_x = \frac{\mathcal{U}_c}{w} , \qquad \mathcal{F}_z = \frac{\mathcal{U}_c}{\ell} = \frac{w}{\ell}\mathcal{F}_x. 
\end{equation}

Next, we substitute the expressions for $\mathcal{F}_z$ from Eq.~\eqref{eq:phi_scale} into Eq.~\eqref{eq:dimless_equilibrium_phix_v1} and multiply by $h_0^2$ to obtain:
\begin{multline}
    \delta^2 D_b\mathcal{F}_x\frac{\partial ^2 \Phi_X}{\partial X^2}+\frac{1}{2}D_b(1-\nu)\mathcal{F}_x\epsilon^2\frac{\partial ^2 \Phi_X}{\partial Z^2}\\
    +\frac{1}{2}\mathcal{F}_x D_b(1+\nu)\epsilon^2\frac{\partial ^2 \Phi_Z}{\partial X \partial Z}
    -\kappa G t h_0^2\mathcal{F}_x\left(\Phi_X +\frac{\partial U}{\partial X}\right) = 0.
    \label{eq:thick_plate_0}
\end{multline}
Under the assumed asymptotic scaling given in Eq.~\eqref{eq:assumption_shallownes}, we retain terms of $\mathcal{O}(\delta^2)$ in the last equation, while dropping the terms of $\mathcal{O}(\epsilon^2)$ in Eq.~\eqref{eq:thick_plate_0}, to obtain:
\begin{equation}
\label{eq:thick_plate_1}
    \delta^2 D_b\frac{\partial ^2 \Phi_X}{\partial X^2}-\kappa G t h_0^2\left(\Phi_X +\frac{\partial U}{\partial X}\right) = 0.
\end{equation}
To balance all terms in the last equation, we must require that $\delta^2D_b \sim \kappa Gt h_0^2$. This scaling can be interpreted in two ways. First, in the ``stiffness space,'' it can be rewritten as
\begin{equation}
\label{eq:scale_thick_plate-1}
    \delta^2 \sim \frac{\kappa G th_0^2}{D_b} =\frac{\text{transverse shear stiffness}}{\text{bending stiffness}},
\end{equation}
which means that the ratio of the transverse shear stiffness to the bending stiffness, though small, is still finite, unlike ``thin-plate'' (Kirchhoff--Love) theory \cite{Love44,TWK59} (referred to as ``classical plate theory'' in \cite{reddy07}), in which it is identically zero. Second, by using the definition of the bending stiffness $D_b$ and the shear modulus $G$ given above, Eq.~\eqref{eq:scale_thick_plate-1} can be rewritten as
\begin{equation}
\label{eq:scale_thick_plate-2}
   6 \kappa{(1-\nu)}\sim (t/w)^2,
\end{equation}
which is an equivalent relation in the ``dimensions space,'' and portrays the relationship between the thickness and the width of the plate in an order of magnitude sense.

Next, Eq.~\eqref{eq:thick_plate_1} can be rewritten as
\begin{equation}
    \label{eq:thick_plate2}
    \mathscr{T} \frac{\partial ^2 \Phi_X}{\partial X^2}-\left(\Phi_X +\frac{\partial U}{\partial X}\right) = 0,\qquad \mathscr{T} := \frac{(t/w)^2}{6\kappa(1-\nu)},
\end{equation}
where a scaled dimensionless thickness $\mathscr{T}$ has been defined for convenience.

Similarly, for Eq.~\eqref{eq:dimless_equilibrium_z_v1}, we substitute $\mathcal{F}_x$ and  $\mathcal{F}_z$ from Eq.~\eqref{eq:phi_scale} and multiply by $h_0^2w^2/D_b$ to obtain:
\begin{multline}
\label{eq:dimless_equilibrium_z_v5}
    \frac{1}{\mathscr{T}}\left(\delta^2\frac{\partial \Phi_X}{\partial X}+\epsilon^2\frac{\partial \Phi_Z}{\partial Z}\right) + \left( \frac{Tw^2}{D_b} + \frac{1}{\mathscr{T}} \right) \left(\delta^2\frac{\partial ^2 U}{\partial X^2 } + \epsilon^2\frac{\partial ^2 U}{\partial Z^2}\right) \\ = - \frac{h_0^2w^2\mathcal{P}_c}{D_b \mathcal{U}_c}P.
\end{multline}
Again, we neglect terms of $\mathcal{O}(\epsilon^2)$ and retain  terms of $\mathcal{O}(\delta^2)$, arriving at
\begin{equation}
\label{eq:dimless_equilibrium_z_v6}
    \frac{1}{\mathscr{T}} \left(\delta^2\frac{\partial \Phi_X}{\partial X}\right) + \left( \lambda + \frac{1}{\mathscr{T}} \right) \left(\delta^2\frac{\partial ^2 U}{\partial X^2 }\right) = - \frac{h_0^2w^2\mathcal{P}_c}{D_b \mathcal{U}_c}P,
\end{equation}
where $\lambda:=Tw^2/D_b$ has been defined as a dimensionless tension-to-bending number. {Although $\lambda<0$ is possible as well (pre-compressed plate), we restrict ourselves to the case of $\lambda>0$ to avoid potentially having to deal with buckled states of the membrane \cite{ZPMSB98}.}

To summarize, the FOSDT equations (in terms of the deformation $U$ and the rotation of the normal $\Phi_X$) for bending of a long and wide plate, initially subject to a uniform isotropic pretension, are
\begin{subequations}\begin{align}
    \label{eq:dimless_equilibrium_z_v7}
        \mathscr{T} \frac{\partial ^2 \Phi_X}{\partial X^2}-\left(\Phi_X +\frac{\partial U}{\partial X}\right) &= 0,\\
    \label{eq:dimless_equilibrium_z_v8}
       \frac{1}{\mathscr{T}} \frac{\partial \Phi_X}{\partial X} + \left( \lambda + \frac{1}{\mathscr{T}} \right) \frac{\partial ^2 U}{\partial X^2 } &= - \frac{h_0^2w^2\mathcal{P}_c}{\delta^2 D_b \mathcal{U}_c}P.
\end{align}\label{eq:dimless_eqs_phi_u}\end{subequations}
The corresponding (four) clamping boundary conditions (BCs) at the channel's lateral sidewalls are
\begin{equation}
 \label{eq:clamping_BCs}
     U|_{X= \pm 1/2} = 0,\qquad \Phi_X|_{X= \pm 1/2} = 0.
\end{equation}
The characteristic deformation scale $\mathcal{U}_c$ remains unknown. It will be determined by considering appropriate balances in Eq.~\eqref{eq:dimless_equilibrium_z_v8}, depending on the order of magnitude of $\lambda$.

\subsection{Solution of the Deformation Equations}
\label{sec:solution_deformation}

It was shown in \cite{CCSS17} (see also Sec.~\ref{sec:fluid} below) that, under the lubrication approximation, the hydrodynamic pressure load can vary \emph{at most} in the flow-wise direction; i.e., $P=P(Z)$ only. Then, the governing differential equations Eqs.~\eqref{eq:dimless_eqs_phi_u} for $\Phi_X$ and $U$ are a set of coupled, inhomogeneous, \emph{ordinary} differential equations (ODEs) in $X$ with constant coefficients. 

Based on the definition of $\lambda$ from Eq.~\eqref{eq:dimless_equilibrium_z_v8}, we can  delineate four regimes of structural deformation:
\begin{itemize}
    \item Regime 1 ($\lambda \ll 1$): Pre-tension is negligible compared to transverse shear and bending, i.e., ${Tw^2}/{D_b} \ll 1$. 
    
    \item Regime 2 ($\lambda = \mathcal{O}(1)$): Pre-tension and transverse shear are comparable to bending: ${T w^2}/{D_b} = \mathcal{O}(1)$, or ${T h_0^2}/{D_b} = \mathcal{O}(\delta^2)$.
    
    \item Regime 3a ($\lambda = \mathcal{O}(1/\delta^2)$): Pre-tension is much stronger that transverse shear and bending: ${T w^2}/{D_b} = \mathcal{O}(1/\delta^2)$, or  ${Th_0^2}/{D_b} = \mathcal{O}(1)$.
    
    \item Regime 3b ($\lambda \gg 1/\delta^2$): Again, pre-tension is much stronger that transverse shear and bending, but there are no longer dominant balances involving $\lambda$ in the governing equation: ${Th_0^2}/{D_b} \gg 1$.
\end{itemize}
Regime 1 is bending dominated, and the problem reduces to the no-pre-tension case considered in previous work \cite{SC18,ADC18}. In Regime 3b, the problem reduces to the trivial case of biaxial stretching of a bar, without any FSI, which is not of interest either. In Regime 2, both pre-tension and bending effects are important. Thus, Regime 2 is of primary interest in this paper, and a solution will be sought for the displacement in this regime. Then, the displacement under Regimes 1 and 3a can be easily found from the solution in Regime 2 as special/limiting cases. A brief independent treatment of Regimes 1 and 3a is presented in {Supplemental Material} Appendix~\ref{app:other_regimes} for completeness.

Next, Eq.~\eqref{eq:dimless_equilibrium_z_v8} can be solved for $\partial\Phi_X/\partial X$: 
\begin{equation}
\label{eq:regime_2_expression}
  \frac{\partial \Phi_X}{\partial X} = -{{\mathscr{T}}} \left[ \frac{w^4\mathcal{P}_c}{D_b \mathcal{U}_c}P+\left( \lambda + \frac{1}{\mathscr{T}} \right) \frac{\partial ^2 U}{\partial X^2 } \right],
\end{equation}
which, in turn, can be differentiated twice with respect to $X$, to obtain:
\begin{equation}
\label{eq:regime_2_expression_2}
  \frac{\partial^3 \Phi_X}{\partial X^3} = - {{\mathscr{T}}} \left( \lambda + \frac{1}{\mathscr{T}} \right) \frac{\partial ^4 U}{\partial X^4 }.
\end{equation}
Taking $\partial/\partial X$ of Eq.~\eqref{eq:dimless_equilibrium_z_v7} and substituting into it the results from Eq.~\eqref{eq:regime_2_expression} and \eqref{eq:regime_2_expression_2}, yields a single ODE for $U$:
\begin{equation}
\label{eq:regime_2_ODE_v1}
    -\left(\lambda\mathscr{T}+1\right)\frac{\partial^4 U}{\partial X^4}+\lambda\frac{\partial^2U}{\partial X^2} +  \frac{w^4\mathcal{P}_c}{D_b \mathcal{U}_c}P =0,
\end{equation}
Thus, to balance all terms (and account for bending, stretching and pressure loading), we must choose the scale of deformation to be
\begin{equation}
\label{eq:deformation_scale}
    \mathcal{U}_c = \frac{w^4\mathcal{P}_c}{D_b}.
\end{equation}

Equation~\eqref{eq:regime_2_ODE_v1} is subject to the four BCs from Eq.~\eqref{eq:clamping_BCs}. Again, two of them need to be converted from BCs on $\Phi_X$ to corresponding BCs on $U$. To that end, differentiate Eq.~\eqref{eq:regime_2_expression} to obtain an expression for ${\partial ^2 \Phi_x}/{\partial X^2}$, which is then substituted into Eq.~\eqref{eq:dimless_equilibrium_z_v7}. Next, evaluate the result at $X=\pm 1/2$ and impose the BCs $\Phi_X|_{X=\pm 1/2} =0$ to obtain the new BCs
\begin{equation}
\label{eq:BC2_Regime2}
\left.\left[\frac{1}{\mathscr{T}}\frac{\partial U}{\partial X} + (\lambda\mathscr{T} + 1)\frac{\partial^3 U}{\partial X^3} \right]\right|_{X=\pm 1/2} = 0.
\end{equation}

By inspection, the particular solution of Eq.~\eqref{eq:regime_2_ODE_v1} is $- \frac{1}{2\lambda}P(Z) X^2$.  For the homogeneous problem, the characteristic polynomial is $\left(\lambda\mathscr{T}+1\right)r^4 - \lambda r^2 = 0$, the roots of which are $r = \left\{\, 0 , \pm \sqrt{\lambda/(\lambda\mathscr{T} + 1)} \,\right\}$, where $r=0$ is a double root. 
Thus, the general solution of Eq.~\eqref{eq:regime_2_ODE_v1} is
\begin{multline}
\label{eq:general_solution_v1}
     U(X,Z) = - \frac{ P(Z)}{2\lambda}X^2 
     + C_1(Z) \exp\left(X\sqrt{\frac{\lambda}{\lambda\mathscr{T}+1}}\right) \\
     + C_2(Z) \exp\left(-X\sqrt{\frac{\lambda}{\lambda\mathscr{T}+1}}\right) 
     + C_3(Z) 
     + C_4(Z) X,
\end{multline}
where $C_{1,2,3,4}(Z)$ are arbitrary functions of integration. 

The $X\mapsto -X$ symmetry of the boundary-value problem (BVP) specified by Eq.~\eqref{eq:regime_2_ODE_v1} and its BCs requires that $C_1(Z) = C_2(Z)$ and $C_4(Z)=0$. Thus, the general solution~\eqref{eq:general_solution_v1} can be rewritten as
\begin{equation}
\label{eq:general_solution_v2}
     U(X,Z) = 2 C_1(Z)\cosh\left(X\sqrt{\frac{1}{\lambda\mathscr{T}+1}}\right)
     + C_3(Z) - \frac{ P(Z)}{2\lambda}X^2,
\end{equation}
The BCs in Eq.~\eqref{eq:BC2_Regime2} require that
\begin{equation}
\label{eq:C_1__Regime2}
    C_1(Z) = \frac{ P(Z)}{4\lambda\sqrt{\lambda(\lambda\mathscr{T}+1)}\sinh \left(\frac{1}{2}\sqrt{\frac{\lambda}{\lambda\mathscr{T}+1}}\right)}.
\end{equation}
Then, the BCs $U|_{X=\pm1/2}=0$ require that
\begin{equation}
\label{eq:C3_Regimen2}
    C_3(Z) = \frac{P(Z)}{2\lambda}\left[\frac{1}{4}-\frac{1}{\sqrt{\lambda(\lambda\mathscr{T}+1)}}\coth\left(\frac{1}{2}\sqrt{\frac{\lambda}{\lambda\mathscr{T}+1}}\right)\right].
\end{equation}

Thus, the complete solution for the cross-sectional deformation profile of the pre-stressed plate is%
\begin{multline}
\label{eq:xsect_sol_FSDT_strech}
   U(X,Z) = \frac{ P(Z)}{2\lambda}\Vast\{\left(\frac{1}{4}-X^2\right) \\ - \left[\frac{\cosh\left(\frac{1}{2}\sqrt{\frac{\lambda}{\lambda\mathscr{T}+1}}\right) - \cosh\left(X\sqrt{\frac{\lambda}{\lambda\mathscr{T}+1}}\right)}{\sqrt{\lambda(\lambda\mathscr{T}+1)}\sinh \left(\frac{1}{2}\sqrt{\frac{\lambda}{\lambda\mathscr{T}+1}}\right)}\right]\Vast\}.
\end{multline}
As a consistency check, we also note that, in the limit of negligible thickness ($\mathscr{T} {\to 0}$), Eq.~\eqref{eq:xsect_sol_FSDT_strech} reduces to the deformation profile of a pre-stressed thin membrane \cite[Eq.~(8)]{ZPMSB98} (see also \cite[Eq.~(10)]{ZYLJYY09}), which is derived independently in {Supplemental Material} Appendix~\ref{app:thin}. 

On the other hand, in the limit $\lambda \to 0$, Eq.~\eqref{eq:xsect_sol_FSDT_strech} reduces to the solution for thick plate without pre-tension, i.e., the solution for Regime 1 from \cite{SC18}, derived independently in {Supplemental Material} Appendix \ref{app:Regime1}. 
In Regime 3a, $\lambda = \mathcal{O}(1/\delta^2)$, i.e., $\lambda \gg 1$, and a straightforward Taylor series expansion of Eq.~\eqref{eq:xsect_sol_FSDT_strech} for $\lambda\to\infty$ gives%
\begin{multline}
\label{eq:xsect_sol_FSDT_strech_regime3_v4}
    U(X,Z) = \frac{P(Z)}{2\lambda} \vast\{ \left(\frac{1}{4}-X^2\right) \\ -  \frac{1}{\lambda} \frac{\cosh \left(\frac{1}{2\sqrt{\mathscr{T}}}\right)-\cosh\left(\frac{X}{\sqrt{\mathscr{T}}}\right)}{\sqrt{\mathscr{T}}\sinh\left(\frac{1}{2\sqrt{\mathscr{T}}}\right) }  
    + \mathcal{O}\left(\frac{1}{\lambda^2}\right) \vast\}.
\end{multline}
The leading-order term in this equation is also derived independently in the {Supplemental Material} Appendix \ref{app:Regime3a}. 

For future reference, Eq.~\eqref{eq:xsect_sol_FSDT_strech} can be put back into its dimensional form:
\begin{multline}
\label{eq:xsect_sol_FSDT_strech_dim}
     u(x,z) = \frac{w^2p(z)}{2T}\Vast\{\left[\frac{1}{4}-\left(\frac{x}{w}\right)^2\right] \\ - \left[\frac{\cosh\left(\frac{1}{2}\sqrt{\frac{\lambda}{\lambda\mathscr{T}+1}}\right) - \cosh\left(\frac{x}{w}\sqrt{\frac{\lambda}{\lambda\mathscr{T}+1}}\right)}{\sqrt{\lambda(\lambda\mathscr{T}+1)}\sinh \left(\frac{1}{2}\sqrt{\frac{\lambda}{\lambda\mathscr{T}+1}}\right)}\right]\Vast\}.
\end{multline}
Then, by $x\mapsto-x$ symmetry, the maximum deformation over the cross-section is its value at $x = 0$:
\begin{multline}
\label{eq:xsect_sol_FSDT_strech_dim_max}
    u_\mathrm{max}(z) = u(0,z) \\ = \frac{w^2p(z)}{2T}\left\{\frac{1}{4} - \frac{\cosh\left(\frac{1}{2}\sqrt{\frac{\lambda}{\lambda\mathscr{T}+1}}\right) -1}{\sqrt{\lambda(\lambda\mathscr{T}+1)}\sinh \left(\frac{1}{2}\sqrt{\frac{\lambda}{\lambda\mathscr{T}+1}}\right)}\right\}.
\end{multline}

\section{Fluid Mechanics}
\label{sec:fluid}

The slenderness of the channel allows us to invoke the lubrication approximation \cite{L07}, according to which the dimensionless velocity field (see \cite{CCSS17} for details) is
\begin{multline}
\label{eq:VZ}
    V_Z(X,Y,Z) = \frac{1}{2}\left(-\frac{\mathrm{d}P}{\mathrm{d}Z}\right)\left[-\frac{t}{2h_0}+\beta U(X,Z)-Y\right] \\ \times \left[Y+\frac{t}{2h_0}+1\right]
\end{multline}
for no-slip boundary conditions at the rigid bottom $Y = -t/(2h_0)-1$ and at the deformed top $Y = -t/(2h_0)+\beta U(X,Z)$ walls. 
The pressure $P$ varies only in the flow-wise $+Z$-direction (meaning, $\mathrm{d}P/\mathrm{d}Z < 0$), thus a complete (not partial) derivative is featured in Eq.~\eqref{eq:VZ}; however, due to FSI, $\mathrm{d}P/\mathrm{d}Z\ne const.$ as it would be in pipe flow \cite{L07}.  
Observe also that in FOSDT, the vertical displacement $U$ does not depend on $Y$ (see Eq.~\eqref{eq:displacement_y} in the Supplemental Material), thus it is the same at $Y=-t/(2h_0)$ (the fluid--solid interface) and at $Y=0$ (the plate's mid-plane).

In Eq.~\eqref{eq:VZ}, $V_Z(X,Y,Z)=v_z(x,y,z)/\mathcal{V}_z$ is the dimensionless velocity in the streamwise direction (recall Fig.~\ref{fig:schematic}), while $U(X,Z)=u_y(x,z)/\mathcal{U}_c$ is the dimensionless deformation of the top wall, as per Eq.~\eqref{eq:scales}. Here, on using Eq.~\eqref{eq:deformation_scale}, we have defined
\begin{equation}
    \label{eq:beta_defined}
    \beta := \frac{\mathcal{U}_c}{h_0} = \frac{w^4\mathcal{P}_c}{D_bh_0}
\end{equation}
as a dimensionless group, which we term the \emph{FSI parameter}. This parameter quantifies the compliance of the plate compared to the characteristic magnitude of the applied hydrodynamic pressure load. 
Then,  the height of the deformed fluid domain is
\begin{equation}
\label{eq:deformed_height}
    H(X,Z) = \frac{h(x,z)}{h_0} = \frac{h_0 + u_y(x,z)}{h_0} 
    = 1 + \beta U(X,Z).
\end{equation}

The dimensionless flow rate is evaluated as the area integral of the streamwise velocity from Eq.~\eqref{eq:VZ}, then written solely in terms of $U(X,Z)$ and $P(Z)$ via Eqs.~\eqref{eq:deformed_height} and \eqref{eq:xsect_sol_FSDT_strech}:
\begin{subequations}\label{eq:flow_ratedefined}\begin{align}
    1 & = \int_{-1/2}^{+1/2}\int_{-\frac{t}{2h_0}-1}^{-\frac{t}{2h_0}-1+H(X,Z)}V_Z(X,Y,Z) \,\mathrm{d}Y\,\mathrm{d}X  \\
    & = -\frac{1}{12}\frac{\mathrm{d}P}{\mathrm{d}Z} \int_{-1/2}^{1/2} H(X,Z)^3 \,\mathrm{d}X \\
    \label{eq:flow_ratedefined3}
    &= -\frac{1}{12}\frac{\mathrm{d}P}{\mathrm{d}Z}\int_{-1/2}^{+1/2}\big[1+3\beta \mathfrak{U}(X)P(Z) + 3\beta^2\mathfrak{U}(X)^2 P(Z)^2\nonumber\\
    &\phantom{-\frac{1}{12}\frac{\mathrm{d}P}{\mathrm{d}Z}\int_{-1/2}^{+1/2}\big[1+} + \beta^3 \mathfrak{U}(X)^3P(Z)^3\big] \mathrm{d}X, 
\end{align}\end{subequations}
where we have introduced the dimensionless deformation-to-pressure ratio
\begin{multline}
\label{eq:mathfrak_u}
    \mathfrak{U}(X) = \frac{U(X,Z)}{P(Z)} = \frac{ 1}{2\lambda}\Vast\{\left(\frac{1}{4}-X^2\right) \\
    - \left[\frac{\cosh\left(\frac{1}{2}\sqrt{\frac{\lambda}{\lambda\mathscr{T}+1}}\right) - \cosh\left(X\sqrt{\frac{\lambda}{\lambda\mathscr{T}+1}}\right)}{\sqrt{\lambda(\lambda\mathscr{T}+1)}\sinh \left(\frac{1}{2}\sqrt{\frac{\lambda}{\lambda\mathscr{T}+1}}\right)}\right]\Vast\}.
\end{multline}
The left-hand side of Eqs.~\eqref{eq:flow_ratedefined} is unity because we have employed a flow-rate-based velocity scale $\mathcal{V}_z = q/(h_0w)$ as in prior work \cite{CCSS17,ADC18}, yielding a dimensionless flow rate $\mathfrak{Q} = q/q = 1$ under steady flow with imposed inlet $q=const$. 

Performing the integration in Eq.~\eqref{eq:flow_ratedefined3} reduces it to a first-order nonlinear ODE in $P(Z)$:
\begin{equation}
\label{eq:flow_rate_p_ODE}
    -12 = \frac{\mathrm{d}P}{\mathrm{d}Z} \left[ 1 + 3\beta \mathcal{I}_{1} P(Z) + 3\beta^2 \mathcal{I}_2 P(Z)^2 + \beta^3 \mathcal{I}_3 P(Z)^3 \right],
\end{equation}
where we have defined
\begin{equation}
\label{eq:mathcal_I}
    \mathcal{I}_{i} := \int_{-1/2}^{+1/2}\mathfrak{U}(X)^i \,\mathrm{d}X.
\end{equation}
Note that $\{ \mathcal{I}_{i} \}_{i=1,2,3}$ are known functions of $\lambda$ and $\mathscr{T}$ (but not $X$ or $Z$), even if obtaining them analytically might be challenging. Now, the ODE~\eqref{eq:flow_rate_p_ODE} is solved subject to the boundary condition that $P(1)=0$ (outlet gauge pressure) to obtain an implicit dimensionless relation for $P(Z)$:
\begin{multline}
\label{eq:flow_rate_p_dimless}
    12(1-Z) = P(Z)\left[1+\frac{3}{2}\beta \mathcal{I}_1 P(Z)+\beta^2\mathcal{I}_2P(Z)^2 \right.\\
    \left. + \frac{1}{4}\beta^3\mathcal{I}_3P(Z)^3\right].
\end{multline}
Finally, the steady flow rate--pressure relation can be put in dimensional form by taking $\mathcal{P}_c = \mathcal{V}_c\mu \ell/h_0^2 = q\mu\ell/(wh_0^3)$ \cite{CCSS17} to be the viscous flow pressure scale for an imposed flow rate:
\begin{multline}
\label{eq:flow_rate_p}
    q = \frac{wh_0^3p(z)}{12\mu(\ell-z)}\Bigg[1+\frac{3}{2}\left(\frac{w^4}{D_bh_0}\right)\mathcal{I}_1 p(z)\\ +\left(\frac{w^4}{D_bh_0}\right)^2\mathcal{I}_2 p(z)^2+\frac{1}{4}\left(\frac{w^4}{D_bh_0}\right)^3\mathcal{I}_3 p(z)^3\Bigg],
\end{multline}
where $\mu$ is the (constant) dynamic viscosity of the Newtonian fluid.

\section{Results} 
\label{sec:computational}

The previous sections were devoted to the derivation of the theory of steady-state fluid--structure interaction (FSI) in a microchannel between the viscous fluid flow within and a pre-stressed elastic top wall clamped on all edges. In this section, we compare the latter theoretical results to direct numerical simulations (DNS) of  FSI performed using the commercial software suite by ANSYS \cite{ANSYS3}. The simulations are two-way coupled to ensure full fidelity. Many of the details of such simulations have been presented in previous publications \cite{CPFY12,CCSS17,SC18,ADC18}. Nevertheless, to ensure that this work is self-contained, a short summary is provided next. 

\subsection{Computational Approach}

ANSYS employs a segregated solution procedure to perform FSI simulations, wherein the mechanical deformation field is solved in the `Static Structural' module, using the finite element method (FEM), while the fluid flow field is solved separately in the `Fluent' module, using the finite volume method (FVM). 

In the Static Structural module, we have switched on the option of `large deformations.' Therefore:
\begin{itemize}
    \item The difference between deformed and undeformed coordinates is maintained.
    \item The logarithmic (Henky) strain and the true (Cauchy) stress are employed as the strain and stress measures, respectively, instead of engineering strain and engineering stress, which would have been employed in a small-strain analysis.
    \item The stiffness matrix in the FEM formulation is a function of the displacements and results in a nonlinear governing equation for each node, which is solved by iterative methods.
\end{itemize} 
Importantly, the assumptions of the plate theory, from which the mathematical model in Sec.~\ref{sec:plates} was derived, are \emph{not} imposed on the numerical solution. Similarly, Fluent solves the steady 3D incompressible Navier--Stokes equation on a deforming domain without any \textit{a priori} approximations. Previously, we carried out mesh refinement studies \cite{SC18}, and we explored choices of algorithms for mesh smoothing \cite{ADC18}, in similar FSI problems. We carry over the lessons learned to the present study to obtain the right blend of numerical accuracy and computational effort.

The distinguishing feature of the FSI simulations carried out in this work is the inclusion of pre-stress in the elastic wall. To this end, we employed two Static Structural modules, instead of one. In the first Static Structural module, forces were imposed on the edges of the structure to induce pre-stress in the elastic wall. The resulting pre-stress distribution was then written to a file. This file containing the information about pre-stress at every node was then read into the second Static Structural module using the `inistate' command. 

The geometric details of the model are given in Table~\ref{Table:Geometry_Parameters_DNS}. The channel has a linearly elastic top wall characterized by a Young's modulus $E = 1.6$ MPa and a Poisson ratio $\nu =0.4999$, similar to PDMS \cite{SC18}. Three values of the uniform pre-tension of the elastic top wall were considered: $T = 13.62$, $27.24$ and $68.11$ N$\cdot$mm, which correspond to $\lambda \equiv Tw^2/D_b =1$, $2$ and $5$, respectively. The remaining three walls of the channel are  rigid. The fluid inside the channel was taken to be water with a constant density $\rho = 997.3$ kg/m$^3$ and dynamic viscosity $\mu = 9.14 \times 10^{-4}$ Pa$\cdot$s. The dimensions of the channel were chosen so that the assumptions of a long and slender geometry, as stated in Eq.~\eqref{eq:assumption_shallownes}, are satisfied, and thus the simulations may be compared to the theory.

\begin{table}
	\centering
	\begin{tabular}{lllllll}
    \hline
	$h_0$ & $w$ & $\ell$ & $t$ & $\delta$ & $\epsilon$ &$t/w$\\ 	
	\hline
    \hline
	0.155 & 1.7 & 15.5 & 0.605 & 0.09 & 0.01 & 0.36\\
    \hline
	\end{tabular}
	\caption{Dimensions and relevant geometric parameters for DNS of FSI in ANSYS. All lengths are given in mm.}
	\label{Table:Geometry_Parameters_DNS}
\end{table}

\subsection{Cross-Sectional Deformation Profile}

\begin{figure}[ht!]
    \centering
    \includegraphics[width=\linewidth]{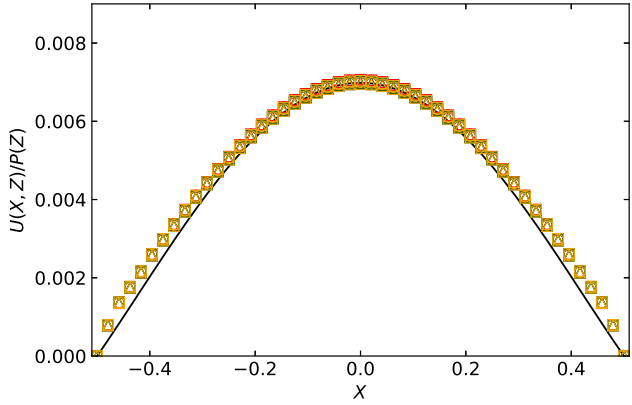}
    \caption{Self-similar collapse of the scaled dimensionless deformation of the channel's elastic plate top wall, across the width of the channel, for different flow rates $q$ and axial locations $z$; $\lambda =2$. The solid curve represents the theoretical prediction from Eq.~\eqref{eq:xsect_sol_FSDT_strech}, the symbols correspond to the results of DNS of FSI in ANSYS.  Colors correspond to flow rates: red is $q=10$ mL/min, yellow is $q=20$  mL/min, gray is $q=30$ mL/min, black is $q=40$ mL/min, green is $q=50$ mL/min, and orange is $q=60$ mL/min. Symbols correspond to different axial locations: $\Box$ is at $z = 4$ mm, $\bigcirc$ is at $z = 8$ mm, and  $\bigtriangleup$ is at $z = 12$ mm. Note that many symbols overlap due to the high quality of the collapse.}
    \label{fig:u_p_dimless_collapse_all}
\end{figure}

A major result of the proposed theory is the self-similar form of the dimensionless cross-sectional deformation profiles scaled by the hydrodynamic pressure, i.e., the ratio $U(X,Z)/P(Z)$ from Eq.~\eqref{eq:xsect_sol_FSDT_strech} is independent of the flow-wise coordinate $Z$. This result connects the local deformation with the local pressure, {forming the theoretical foundation for the hydrostatic bulge test}. To verify this result of the theory, in Fig.~\ref{fig:u_p_dimless_collapse_all} we plot the results from ANSYS FSI simulations for $U(X,Z)/P(Z)$ as a function of $X$ alongside the prediction from  from Eq.~\eqref{eq:xsect_sol_FSDT_strech}. We observe that the simulations (represented by symbols and colors, corresponding to the different flow rates $q$ and evaluated at different flow-wise locations $z$) collapse neatly onto a single curve, which closely matches the theoretical profile (solid curve). In practice, validation of this prediction cannot be carried out in a noninvasive manner due to the need to measure the deformation at several flow-wise locations.

\begin{figure}
    \centering
    \includegraphics[width=\linewidth]{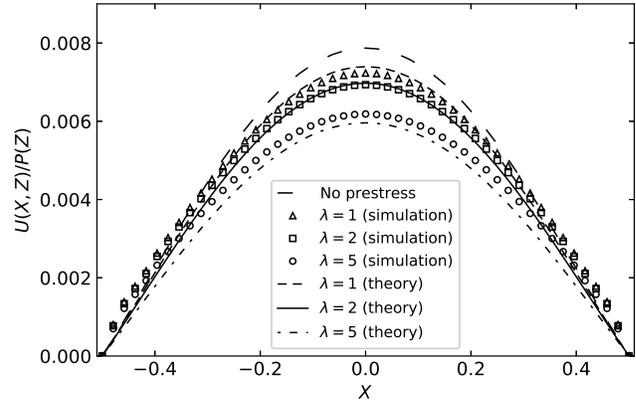}
    \caption{Self-similar scaled dimensionless deformation of the elastic plate top wall of the channel, for different bending-to-tension ratios $\lambda$, at an inlet flow rate of $q =30$ mL/min. The curves represents the theoretical prediction from Eq.~\eqref{eq:xsect_sol_FSDT_strech}, while the symbols correspond to the results of DNS of FSI in ANSYS.}
    \label{fig:u_p_dimless_collapse}
\end{figure}

Next, we carried out the simulations for fixed $q$ but different values of the pre-tension $T$. In Fig.~\ref{fig:u_p_dimless_collapse}, the profile $U(X,Z)/P(Z)$ from ANSYS simulation is compared to the theoretical profile from Eq.~\eqref{eq:xsect_sol_FSDT_strech}, for different values of the bending-to-tension ratio $\lambda$. We observe good match between the theoretical prediction and the results of simulation, but it worsens as $\lambda$ increases. A possible explanation may be that significant stretching occurs in the structure at high $\lambda$, which invalidates the small strain assumption employed in the FOSDT of plates. Also, the agreement is better at the center of the cross-section, compared to the sides, 
{which can be attributed to the use of clamped boundary conditions in the theory, while a 3D zero-displacement boundary condition for the nodes along the $x=\pm w/2$ planes is imposed in simulations. Others have used the so-called ``elastically clamped'' boundary conditions \cite{ZYSLYL08, ZYLJYY09} to improve the agreement near the edges. 
However, the elastically clamped boundary conditions involve a free parameter, whose value must be determined from additional numerical simulation of the particular plate geometry \cite{PB99}. Therefore, the use of elastically clamped boundary conditions pose their own set of challenges, while yielding at a best a modest improvement in the already quite good match between the theoretical and simulated deformation profiles.}

\subsection{Flow Rate--Pressure Drop Relationship}

Next, we shift our focus to the flow rate--pressure drop relationship obtained in closed-form from the theory as Eq.~\eqref{eq:flow_rate_p}. This result involves variables that need to be measured only at the inlet and/or the outlet of the microchannel; these measurements can be done noninvasively. Therefore, there is no need to measure quantities inside the system (channel) to obtain an estimate of the material properties from Eq.~\eqref{eq:flow_rate_p}. The latter idea underpins the proposed hydrodynamic bulge test, which renders the system a ``black box'' for experimental materials characterization, unlike the hydrostatic bulge test, which requires measuring $u_\mathrm{max}(z)$ at some axial position $z$ and inverting Eq.~\eqref{eq:xsect_sol_FSDT_strech_dim_max} to determine $E$ (via $D_b$ in $\lambda$).

\begin{figure}
    \centering
    \includegraphics[width=\linewidth]{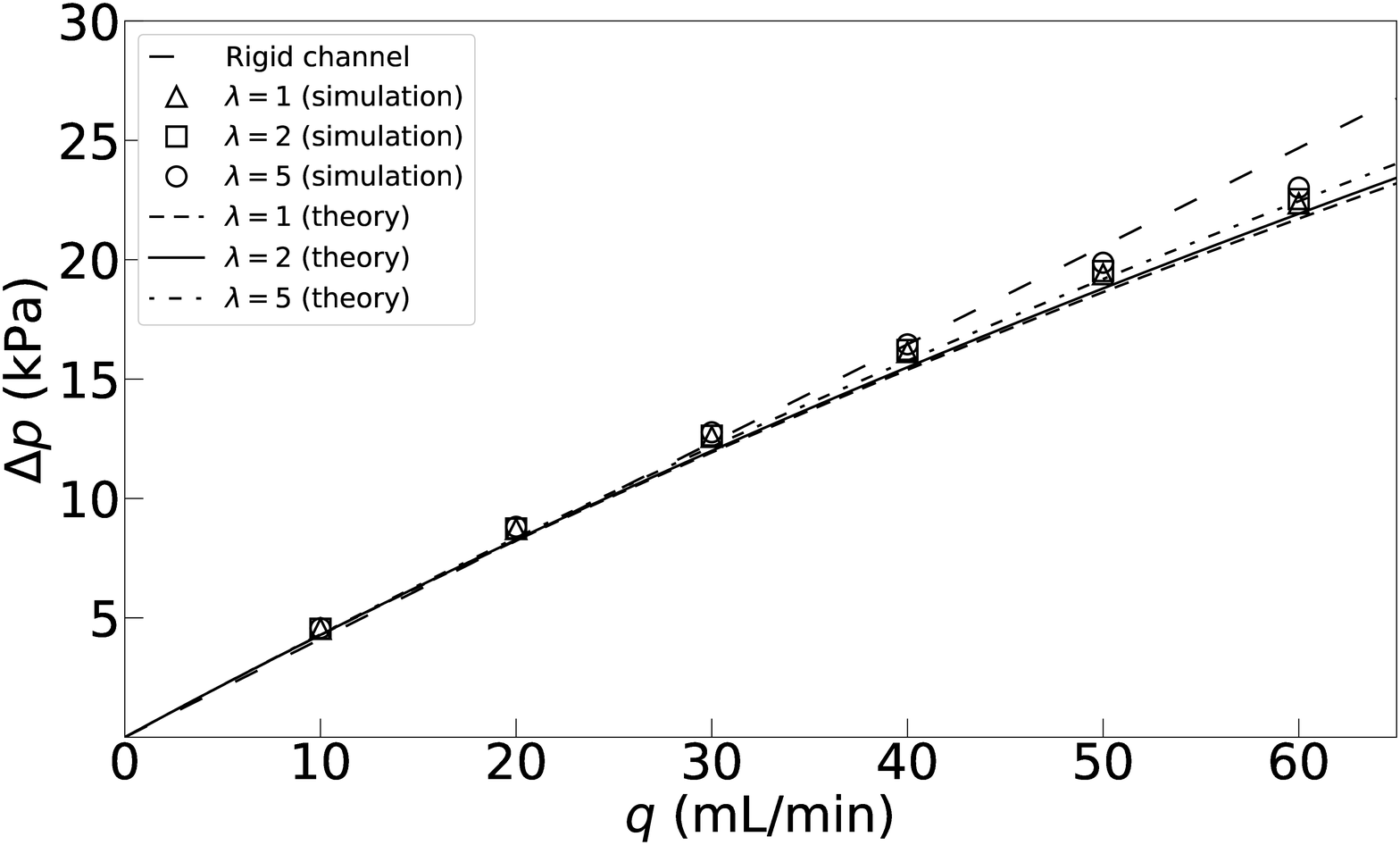}
    \caption{Comparison of flow rate--pressure drop relationship from numerical simulation (markers) and the theoretical dependence found from inverting Eq.~\eqref{eq:flow_rate_p} (solid curves), for different values of the tension-to-bending ratio $\lambda$ to highlight the effect of pre-tension in the plate.}
    \label{fig:q_dp_dimensional}
\end{figure}

To illustrate our FSI theory, in Fig.~\ref{fig:q_dp_dimensional}, we plot the full pressure drop $\Delta p = p(0)-p(\ell)=p(0)$, as calculated from Eq.~\eqref{eq:flow_rate_p}, as a function of the volumetric flow rate $q$, for different tension-to-bending ratios $\lambda$. Additionally, the corresponding results from our ANSYS FSI simulations are shown as symbols. Clearly, the theory agrees with the simulations for the range of $q$ and $\lambda$ considered. An increase in $\lambda$ causes the pressure drop to increase, because of the decrease in deformation as pre-tension ``stiffens'' the plate. The match worsens at larger $q$ and $\lambda$ due to ``stronger'' FSI. For each $\lambda$, the maximum error between theory and simulation occurs at the maximum flow rate $q = 60$ mL/min; still, this maximum relative error is just $\approx 2.53\%$ for $\lambda =5$.

\subsection{Characterization of Material Properties and Range of Validity of the Theory}

The goal of this work is to introduce a theory of hydrodynamic bulge testing, wherein the material properties of a finite-thickness elastic membrane (plate) are characterized using a pressure drop measurement and the relationship in  Eq.~\eqref{eq:flow_rate_p}, without measuring the membrane's transverse deformation. To achieve this goal, the measured pressure drop, the imposed flow rate, and the known geometric dimensions are substituted into Eq.~\eqref{eq:flow_rate_p}, which is then solved using the bisection method \cite[Ch.~5]{CC15} in a Python script using SciPy \cite{SciPy}, to obtain the Young's modulus $E$ given the Poisson ratio $\nu$.
 
\begin{figure}
    \centering
    \includegraphics[width=\linewidth]{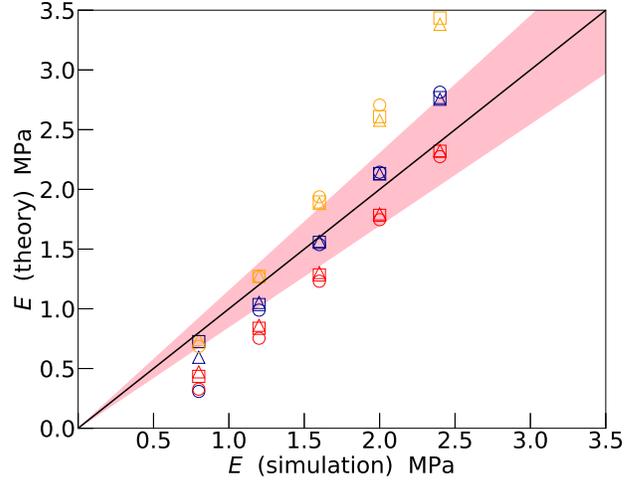}
    \caption{Estimation of the elastic modulus $E$ from a hydrodynamic bulge test versus its ``true'' value (used in the simulations). Colors correspond to different flow rates: orange is $q = 60$ mL/min, blue is $q = 80$ mL/min, and red is $q =100$ mL/min. Symbols correspond to different values of the tension-to-bending ratio $\lambda$: $\bigcirc$ is $\lambda =5$, $\Box$ is $\lambda  = 2$, and $\bigtriangleup$ is $\lambda = 1$. All  other quantities are as given in Table~\ref{Table:Geometry_Parameters_DNS}. The shaded area corresponds to an uncertainty of $\pm 15\%$ in $E$.}
    \label{fig:E_p_comparison}
\end{figure}

To demonstrate how a hydrodynamic bulge test might work, we carried out FSI simulations for different values of  the elastic modulus $E$ of the top wall. The simulations were carried out for flow rates of $q = 60$ mL/min, $80$ mL/min, and $100$ mL/min for $\lambda =1$, $2$, and $5$. The resulting pressure drop from the simulation was used to predict the value of $E$ by inverting Eq.~\eqref{eq:flow_rate_p}. The results of this analysis are shown in Fig.~\ref{fig:E_p_comparison}.

Clearly, there is acceptable match between the actual (here, simulation) values and the estimated (here, theoretical) values of $E$, for the chosen range that is typical of PDMS. The quality of this match is gauged by the closeness of the symbols to the line with slope $1$ passing through the origin. The maximum error is about $43 \%$  for the case of $q = 60$ mL/min and $E = 2.4$ MPa. We note that the match is better for stiffer walls (larger values of $E$) at higher flow rates, and for softer walls (smaller values of $E$) at lower flow rates. 

This observation can be explained by considering the regime of validity of our theory, which is given in mathematical terms as:
\begin{equation}
\label{eq:inequality_1}
    \{ t \lesssim w \sim \mathcal{U}_c \} \ll \ell,
\end{equation}
where
\begin{equation}
    \mathcal{U}_c = \frac{w^4\mathcal{P}_c}{D_b} =\frac{q\mu\ell w^3}{D_b h_0^3}=\frac{12q\mu\ell(1-\nu^2) w^3}{E t^3 h_0^3}.
\end{equation}
Here, the inequality $t \lesssim w$ means that we have accounted for moderate (rather than vanishing, $t \ll w$) plate thickness. The scaling $\mathcal{U}_c \sim w$ means that we have accounted for moderate rotations in the equilibrium equations, by inclusion of $\mathcal{N}$ and, thus, the plate's bending response is coupled with its stretching response. The inequality of $w \ll \ell$ is necessary to ensure that the lubrication approximation is valid (for the fluid mechanics problem), so that cross-section deformation profiles are decoupled from each other in the flow-wise direction (for the structural mechanics problem). Therefore, if the characteristic deformation $\mathcal{U}_c$ is large compared to the dimensions of the channels, i.e., $\mathcal{U}_c > w$, the structural mechanics problem is no longer linear and our FSI theory breaks down. On the other hand, if $\mathcal{U}_c$ is extremely small, i.e., $\mathcal{U}_c \ll w$, and the FSI in the system is ``weak,'' the estimate of $E$ deteriorates. In the limiting case of a rigid channel, in which there is obviously no FSI, it would not be possible to estimate $E$ at all because there is no deformation.

\begin{figure}
    \centering
\resizebox{\columnwidth}{!}{
\begin{tikzpicture}[node distance=1.25cm]
\node (1) [process2, fill=blue!15] {Fix geometry,  viscosity and Poisson ratio: $h_0$, $w$, $\ell$, $t$, $\mu$, $\nu$.};
\node (2) [process2, below of=1, fill=blue!15] {Guess $E_\mathrm{estimate}$ ($\simeq1$MPa for PDMS).};
\node (3) [process2, below of=2, fill=green!20] {Calculate a suitable $q$ for the experiment from Eq.~\eqref{eq:validity_range}.};
\node (4) [process2, below of=3, fill=green!20] {Conduct experiment and measure $\Delta p$.};
\node (5) [process2, below of=4, fill=green!20] {Calculate new $E_\mathrm{estimate}$ from Eq.~\eqref{eq:flow_rate_p} using measured $\Delta p$.};
\node (6) [decision, right of=4, xshift=4.5cm, yshift=0cm, fill=orange!30] {Satisfy Eq.~\eqref{eq:validity_range}?};
\node (7) [end, below of=5, fill=red!40] {Stop. $E_\mathrm{estimate}$ obtained.};
\draw [arrow] (1) -- (2);
\draw [arrow] (2) -- (3);
\draw [arrow] (3) -- (4);
\draw [arrow] (4) -- (5);
\draw [arrow] (5.east) -- ++(0.3,0) -- ++(0,0.75) -- ++(-1.25,0) |- (6.west); 
\draw [arrow] (6) |- node[anchor=west]{No}(3);
\draw [arrow] (6) |- node[anchor=west]{Yes}(7);
\end{tikzpicture}
}
    \caption{Flow chart of how to iteratively apply the hydrodynamic bulge testing methodology to estimate the Young's modulus $E$ of a plate, starting from a guess.}
    \label{fig:iteration}
\end{figure}
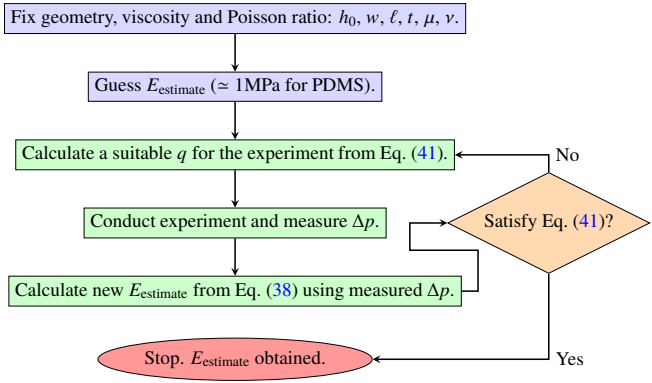

{To quantify the above-identified requirement of ``sufficient FSI'' via  deformation that is still in the linearly elastic regime, we can employ Eq.~\eqref{eq:inequality_1} to restrict the range of $\mathcal{U}_c/{w}$ values for which the hydrodynamic bulge test is expected to be accurate. Observing that $\beta\delta=\mathcal{U}_c/{w}$, it is more convenient to write this restriction as one on the FSI parameter $\beta$ introduced in Eq.~\eqref{eq:beta_defined}. Based on applying the hydrodynamic bulge test idea to simulation data from the present study, as well as previous simulations \cite{SC18} and experiments \cite{MY19} without pre-stress, we suggest the order-of-magnitude guideline:
\begin{equation}
    1 \lesssim \beta \equiv \frac{12q\mu\ell (1-\nu^2)w^3}{Et^3 h_0^4} \lesssim 10,
\label{eq:validity_range}
\end{equation}
where the upper range of values is suitable for non-pre-stressed plates ($\lambda=0$), while the lower range of values should be preferred in the case of a plate stiffened by pre-stress ($\lambda=\mathcal{O}(1)$).} This guideline is an important result in practice. Since one can control $q$, $\mu$, $\ell$, $t$, $w$ and $h_0$, then it always possible to set up a sample, to be characterized by the proposed hydrodynamic bulge test, such that the bulge test is accurate. {However, since Eq.~\eqref{eq:validity_range} already contains $E$, it must be applied in a recursive manner to design the hydrodynamic bulge test experiment, as show in Fig.~\ref{fig:iteration}. Importantly, the iteration process only requires updating the flow rate $q$ in the experiment (easily controlled by a pump), thus it does \emph{not} require modification of the microchannel geometry, once it is manufactured.}

\section{Error Propagation and Sensitivity Analysis}
\label{sec:sensitivity}

It is important to compare the hydrostatic and hydrodynamic bulge tests with respect to error propagation via a sensitivity analysis. Due to measurement errors \cite{M84}, any experimental observation has an uncertainty associated with it. The uncertainty in the independent (measured) variable propagates to an uncertainty in the dependent (estimated) variable. In this section, we discuss examples of how errors propagate in the hydrostatic and the hydrodynamic bulge testing theories. Specifically, we simulate how uncertainty in the corresponding independent variables ($\Delta p$ in the hydrodynamic case and $u_\mathrm{max}$ in the hydrostatic case) leads to an uncertainty in the dependent variable, the Young's modulus $E$, and compare the two cases to each other.

In our theory from Sec.~\ref{sec:fluid}, the relationship between the dependent variable and the independent variables is given by Eq.~\eqref{eq:flow_rate_p} in conjunction with Eqs.~\eqref{eq:mathfrak_u} and \eqref{eq:mathcal_I}. These set of coupled equations is not amenable to a standard uncertainty quantification by analytical means, such as a Taylor series-based root-mean-squared error propagation \cite[Sec.~4.7]{M84}. This situation is unlike the hydrostatic case in which $E$ is determined (via $D_b$ in $\lambda$) by measuring $u_\mathrm{max}(z)$ at some axial position $z$ and inverting Eq.~\eqref{eq:xsect_sol_FSDT_strech_dim_max}, which is amenable to an error propagation analysis. Hence, we take a statistical approach and perform Monte Carlo simulations of error propagation. 

\begin{figure}
    \centering
    \includegraphics[width=\linewidth]{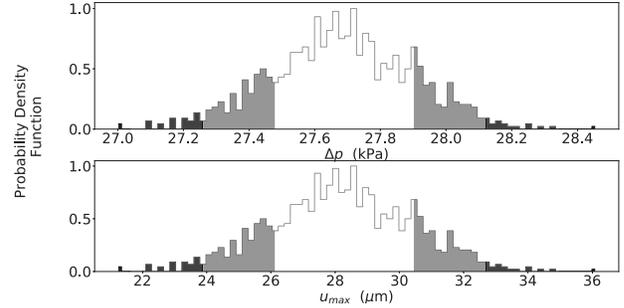}
    \caption{Realizations of the noisy distributions of the total pressure drop $\Delta p$ (top) and the maximum deformation $u_\mathrm{max}$ (bottom) as an input to the Monte Carlo simulation for error propagation under the hydrostatic and hydrodynamic bulge test, respectively. Each shaded band has a width of one standard deviation.}
    \label{fig:Monte_Carlo_input_1}
\end{figure}

The Monte Carlo simulation of error propagation is straightforward. The independent variables $u_\mathrm{max}$ and $\Delta p$ are replaced by random variables, which are sampled from a normal distribution. The normal distribution is, in turn, determined from a nominal value, the given mean $\overline{u_\mathrm{max}}$ or $\overline{\Delta p}$ in Table~\ref{table:MonteCarloInput}, and the upper and lower limits on uncertainty $\pm\Upsilon$ as a percentage, from which the standard deviations $\varsigma$ of the distributions are
\begin{equation}\label{eqs:st_dev}
    \varsigma_{u_\mathrm{max}} = \left[\frac{\Upsilon_{u_\mathrm{max}}}{F^{-1}(0.9)}\right]\overline{u_\mathrm{max}},\qquad
    \varsigma_{\Delta p} = \left[\frac{\Upsilon_{\Delta p}}{F^{-1}(0.9)}\right]\overline{\Delta p}.
\end{equation}
The factor $\Upsilon/F^{-1}(0.9)$ in Eqs.~\eqref{eqs:st_dev}, where $F^{-1}(0.9)$ is the inverse of the normal cumulative density function at the 90th quantile,  ensures that $90\%$ of the area under probability density function is below the specified upper limit ($+\Upsilon\%$) and similarly $90\%$ of the area under probability density function is above the specified lower limit ($-\Upsilon\%$). A total of $1000$ samples were taken of the input random variables, as a trade-off between the computational effort expended and the desired accuracy of probabilistic models. Example distributions of the input variables are shown in Fig.~\ref{fig:Monte_Carlo_input_1}.

\begin{table}
\small
\centering
\begin{tabular}{@{}llll@{}}
\toprule
Bulge Test & Input & Mean $\overline{(\cdot)}$ & Uncertainty $\Upsilon$ \\ \midrule\midrule
Hydrostatic & $u_\mathrm{max}$ & 28.296 $\mu$m & $\pm10\%$ \\ \midrule
Hydrodynamic & $\Delta p$ &  27.691 kPa& $\pm1\%$ \\ \bottomrule
\end{tabular}
\caption{Statistics of the input distribution for the Monte Carlo simulation of error propagation under the hydrostatic and hydrodynamic bulge tests. The mean values of the independent variables are the ones used in the ANSYS simulation for $q = 80$ mL/min, $\lambda =2$ and $E = 1.60$ MPa.}
\label{table:MonteCarloInput}
\end{table}

The means match the deterministic values used in the ANSYS simulations corresponding to a flow rate $q= 80$ mL/min, a tension-to-bending ratio $\lambda = 2$, and Young's modulus $E = 1.6$ MPa. The uncertainty in the pressure drop measurement corresponds to that of a  standard off-the-shelf pressure measurement device like Omega PXM409-007BDWUI. On the other hand, the uncertainty in the deformation has been taken to be an order of magnitude larger, at $10\%$, which is close to the uncertainty in deformation measurements extracted from the experimental data in \cite{Hetal18, RS16}.

\begin{figure}
    \centering
    \includegraphics[width=\linewidth]{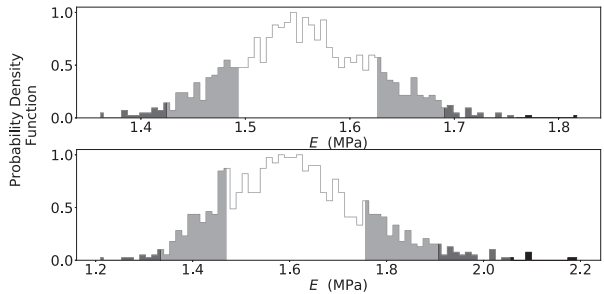}
    \caption{Distribution of the estimate of $E$ obtained from the Monte Carlo simulation of the hydrodynamic (top) and hydrostatic (bottom) bulge test. Each shaded band has a width of one standard deviation.}
    \label{fig:MonteCarlo_Output_1}
\end{figure}

The mean and the standard deviation of the output samples for $E$ were computed for both  models. The results are shown in Fig.~\ref{fig:MonteCarlo_Output_1} and in Table~\ref{table:MonteCarloOutput}. To gauge the sensitivity of the estimate of $E$ to the input distributions of $u_\mathrm{max}$ and $\Delta p$, we employed a statistical rank order correlation, specifically Kendall's tau rank correlation coefficient, available in Python's SciPy module \cite{SciPy}. The value of Kendall's tau rank correlation lies between $-1$ and $+1$; a value of $+1$ denotes strong positive correlation, a value of $-1$ denotes strong negative correlation, while a value of $0$ denotes no correlation at all. From Fig.~\ref{fig:MonteCarlo_Output_1} and Table~\ref{table:MonteCarloOutput}, we conclude that estimates of elastic modulus  obtained from the hydrostatic bulge test are more accurate compared to those obtained from the hydrodynamic bulge test, though the difference is not very large ($\approx 2.5\%$).  However, the noise in the estimates $E$ is much larger for the hydrostatic bulge test than for the hydrodynamic bulge test, as evidenced by the larger standard deviation of the hydrostatic bulge test's output distribution. The higher noise in the estimated variable $E$ is attributed to the higher noise in the measured variable $u_\mathrm{max}$, as the absolute value of the rank correlation coefficient (Kendall's $\tau$) is approximately the same for both models.

\begin{table}
\small
\centering
\begin{tabular}{@{}lllll@{}}
\toprule
Bulge Test & \begin{tabular}[c]{@{}c@{}} $\bar{E}$ (MPa)\end{tabular} & $\varsigma_E$ (MPa) & $\varsigma_E/\bar{E}$ & Kendall's $\tau$ \\ \midrule\midrule
Hydrostatic & $1.61$ & 0.14& $0.087$ & $-0.99$ \\ \midrule
Hydrodynamic & $1.57$ &  0.07& $0.045$ &  $1.0$\\ \bottomrule
\end{tabular}
\caption{Statistics for the estimate of the Young's modulus $E$ obtained from the Monte Carlo simulation of the hydrostatic and hydrodynamic bulge tests; $q = 80$ mL/min and $\lambda=2$. $\bar{E}$ is the mean, and $\varsigma_E$ is the standard deviation.}
\label{table:MonteCarloOutput}
\end{table}

\section{Conclusion}
\label{sec:conclusion}

In this paper, we proposed a hydrodynamic bulge testing technique for soft materials characterization problems relevant to design of microfluidic systems and devices. Specifically, we derived a theory of the fluid--structure interaction (FSI) between a pre-stressed linearly elastic plate with finite thickness and a viscous fluid flow underneath it. The flow rate--pressure drop relationship for the case when the elastic plate is clamped as the top wall of a rectangular microchannel conveying a ``slow'' viscous Newtonian fluid flow (low Reynolds number), was obtained in the form of Eq.~\eqref{eq:flow_rate_p}. Then, we showed that this relationship can be inverted numerically to characterize the material properties of the elastic plate, specifically its Young's modulus $E$, by only measuring the total pressure drop across its length. We argued that, in microfluidics, a measurement of the pressure drop is easier and/or more accurate than a measurement of the membrane deflection due to the (hydrodynamic) pressure of the flow underneath it. 

We also carried out three-dimensional direct numerical simulations of fluid--structure interactions using the commercial computational engineering platform by ANSYS. These simulations did not require any of the assumptions used to derive the mathematical model. The deformation profile and pressure drop obtained from the simulations showed favorable agreement with the predictions of our model, thus validating it. 

Overall, from the mechanics point of view, pre-stressing the membrane makes it appear ``stiffer,'' thus its deformation (induced by either hydrodynamic or hydrostatic pressure) is smaller than a corresponding initially stress-free plate. To sustain the same flow rate in a microchannel with an initially pre-stressed top wall thus requires a larger pressure drop. These conclusions were drawn from the general displacement profile, Eq.~\eqref{eq:xsect_sol_FSDT_strech}, which we believe is a novel result because the expression accounts for the non-negligible thickness of the membrane ($t/w\not\to0$), while the current literature on bulge testing concerns  thin-film membranes ($t/w\to0$) \cite[Eq.~(8)]{ZPMSB98}.

Next, the simulations were used as ``ground truth'' (in lieu of experiments) to establish the predictive power of hydrodynamic bulge tests. Specifically, a region in the parameter space was identified and represented as Eq.~\eqref{eq:inequality_1}. Through  Fig.~\ref{fig:E_p_comparison}, this parameter space region of validity was quantified {Eq.~\eqref{eq:validity_range} was proposed as a guideline to the experimentalist for obtaining accurate results from the hydrodynamic bulge test.} Furthermore, a sensitivity analysis, performed through Monte Carlo simulation, showed that the hydrodynamic bulge test's estimate is only slightly less precise than a hydrostatic bulge test, while allowing a greater degree of ``control'' over error propagation.

{Due to the long and shallow channel geometries encountered in PDMS-based microfluidics, the strains encountered in the elastic wall are small enough to justify the use of the linear theory of elasticity. For example, for $q = 50$ mL/min at $\lambda =2$, the maximum normal strain according to our ANSYS simulation is $3\%$, which acceptable under the small-strain assumption made in our theory. At larger strains, it is know that PDMS may exhibit a hyperelastic response \cite{KKJ11}. Thus, in future work, it would be of interest to extend the proposed theory to capture this nonlinear material behavior.} The proposed hydrodynamic bulge testing technique could also be extended to handle liquid blister tests \cite{CVB08}, which are used to measure the strength of bonding (via the work of adhesion), if the fluid layer is made much thinner than the solid film. Beyond bulge tests, the FSI between a viscous fluid and a pre-stressed plate-like elastic structure can be harnessed to create soft microfluidic actuators \cite{BGB18}. Similar multiphysics problems can also be motivated by biomedical and physiological applications, such as the reopening of strongly collapsed airways \cite{DHTJ17}. These problems are unsteady \cite{MCSPS19}, thus one must obtain dynamic equations for the motion of the fluid front during expansion (or collapse) \cite{HBDB14,MHT15,EG16,BN18}. Therefore, the present analysis could be extended/become the foundation of further research on these problem as well.

\section*{Acknowledgements}
V.A.\ and I.C.C.\ were supported, in part, by the U.S.\ National Science Foundation under grant No.\ CBET-1705637. S.C.M.\ was supported by the 2019 Purdue Undergraduate Research Experience (PURE) under the Purdue-India Initiative. 

\clearpage

\section*{Nomenclature}
\setlength\tabcolsep{2pt} 
\begin{supertabular}{rcl}
$C_i$ &$=$ & constants of integration\\
$D_s$ &$=$ & extensional rigidity of the plate, Pa$\cdot$m\\
$D_b$ &$=$ & bending rigidity of the plate, Pa$\cdot$m$^3$\\
$E$ &$=$ & Young's modulus of the plate, Pa\\
$F^{-1}$ &$=$ & inverse of the cumulative distribution function \\
$\mathcal{F}_x$, $\mathcal{F}_z$ &$=$ & characteristic scales of the rotations of the normal  \\
$G$ &$=$ & shear modulus, Pa\\
$h_0$ &$=$ & height of the undeformed channel, m\\
$H$ &$=$ & dimensionless height of the deformed channel \\
$\mathcal{I}_{i}$&$=$ & integral of $i$th power of the ratio of dimensionless \\
& & deformation to dimensionless pressure\\
$\ell$ &$=$ & length of channel, m\\
$M$ &$=$ & bending moment, N$\cdot$ m\\
$N$ &$=$ & normal stress resultant, Pa$\cdot$m\\
$\mathcal{N}$ &$=$ & coupling term between bending  and stretching \\
& & in the equation of equilibrium, Pa\\
$p$ &$=$ & pressure, Pa\\
$P$ &$=$ & dimensionless pressure \\
$\mathcal{P}_c$ &$=$ & characteristic pressure scale, Pa\\
$q$ &$=$ & flow rate, mL/min \\
$Q$ &$=$ & shear stress resultant, Pa$\cdot$m\\
$t$ &$=$ & thickness of the plate, m\\
$T$ &$=$ & pre-tension, Pa$\cdot$m \\
$\mathscr{T}$ &$=$ & dimensionless thickness of the plate  \\
$u$ &$=$ & deformation, m\\
$U$&$=$ & dimensionless deformation \\
$\mathcal{U}_c$ &$=$ & characteristic deformation scale, m\\
$\mathfrak{U}$ &$=$ & ratio of dimensionless deformation \\
& & to dimensionless pressure\\
$v$ &$=$ & fluid velocity, m/s\\
$V$ &$=$  & dimensionless fluid velocity \\
$\mathcal{V}_c$ &$=$  & characteristic velocity scale, m/s\\
$w$ &$=$ & width of channel, m\\
$x,y,z$ &$=$ & rectangular coordinates, m \\
$X,Y,Z$&$=$ & dimensionless rectangular coordinates \\
&&\textbf{Greek Symbols}\\
$\epsilon$&$=$ & ratio of channel height and length, $h_0/\ell$  \\
$\delta$&$=$ & ratio of channel height and width, $h_0/w$  \\
$\Upsilon$ &$=$ & uncertainty  \\
$\tau$&$=$ & Kendall's tau correlation  \\
$\varsigma$&$=$ &standard deviation \\
$\phi$ &$=$ & rotation of the normal\\
$\Phi$ &$=$ & dimensionless rotation of normal\\
$\doubleunderline{\varPhi}$&$=$ & curvature tensor\\
$\gamma$ &$=$ & strain\\
$\sigma$ &$=$ & stress,   Pa\\
$\kappa$ &$=$ & Timoshenko's shear correction factor\\
$\lambda$ &$=$ & ratio of pre-tension to bending rigidity\\
$\beta$&$=$ & fluid--structure interaction parameter \\
$\nu$ &$=$ & Poisson ratio \\
$\mu$&$=$ & viscosity of the fluid, Pa$\cdot$s \\
$\Delta p$&$=$ & pressure drop, Pa \\
&&\textbf{Subscripts}\\
$x,y,z$ &$=$ & flow or deformation direction, dimensional \\
$X,Y,Z$&$=$ & flow or deformation direction, dimensionless \\
$0$ &$=$ &along the plate's mid-plane \\
$c$ &$=$ &characteristic \\
$i$ &$=$ &positive integer \\
$\mathrm{max}$ &$=$ & maximum \\
\end{supertabular}


\balance

\bibliographystyle{elsarticle-num.bst}
{\footnotesize\bibliography{references2.bib}}

\begin{thebibliography}{10}
\expandafter\ifx\csname url\endcsname\relax
  \def\url#1{\texttt{#1}}\fi
\expandafter\ifx\csname urlprefix\endcsname\relax\def\urlprefix{URL }\fi
\expandafter\ifx\csname href\endcsname\relax
  \def\href#1#2{#2} \def\path#1{#1}\fi

\bibitem{VN92}
J.~J. Vlassak, W.~D. Nix, {A new bulge test technique for the determination of
  Young's modulus and Poisson's ratio of thin films}, J. Mat. Res. 7 (1992)
  3242--3249.
\newblock \href {http://dx.doi.org/10.1557/JMR.1992.3242}
  {\path{doi:10.1557/JMR.1992.3242}}.

\bibitem{SN92}
M.~K. Small, W.~D. Nix, {Analysis of the accuracy of the bulge test in
  determining the mechanical properties of thin films}, J. Mat. Res. 7 (1992)
  1553--1563.
\newblock \href {http://dx.doi.org/10.1557/JMR.1992.1553}
  {\path{doi:10.1557/JMR.1992.1553}}.

\bibitem{MW02}
J.~C. McDonald, G.~M. Whitesides, {Poly(dimethylsiloxane) as a material for
  fabricating microfluidic devices}, Acc. Chem. Res. 35 (2002) 491--499.
\newblock \href {http://dx.doi.org/10.1021/ar010110q}
  {\path{doi:10.1021/ar010110q}}.

\bibitem{ALA99}
D.~Armani, C.~Liu, N.~Aluru, {Re-configurable fluid circuits by PDMS elastomer
  micromachining}, in: Twelfth IEEE International Conference on Micro Electro
  Mechanical Systems (MEMS'99), 1999, pp. 222--227.
\newblock \href {http://dx.doi.org/10.1109/MEMSYS.1999.746817}
  {\path{doi:10.1109/MEMSYS.1999.746817}}.

\bibitem{ACJCDWWW00}
J.~R. Anderson, D.~T. Chiu, R.~J. Jackman, O.~Cherniavskaya, J.~C. McDonald,
  H.~Wu, S.~H. Whitesides, G.~M. Whitesides, {Fabrication of topologically
  complex three-dimensional microfluidic systems in PDMS by rapid prototyping},
  Anal. Chem. 72 (2000) 3158--3164.
\newblock \href {http://dx.doi.org/10.1021/ac9912294}
  {\path{doi:10.1021/ac9912294}}.

\bibitem{XW98}
Y.~Xia, G.~M. Whitesides, {Soft lithography}, Annu. Rev. Mater. Sci. 28 (1998)
  153--184.
\newblock \href {http://dx.doi.org/10.1146/annurev.matsci.28.1.153}
  {\path{doi:10.1146/annurev.matsci.28.1.153}}.

\bibitem{JMTT14}
I.~D. Johnston, D.~K. McCluskey, C.~K.~L. Tan, M.~C. Tracey, {Mechanical
  characterization of bulk Sylgard 184 for microfluidics and microengineering},
  J. Micromech. Microeng. 24 (2014) 35017.
\newblock \href {http://dx.doi.org/10.1088/0960-1317/24/3/035017}
  {\path{doi:10.1088/0960-1317/24/3/035017}}.

\bibitem{J08}
W.~P. Jackson,
  \href{https://thesis.library.caltech.edu/2322/}{{Characterization of Soft
  Polymers and Gels using the Pressure-Bulge Technique}}, Ph.D. thesis,
  California Institute of Technology (2008).
\newline\urlprefix\url{https://thesis.library.caltech.edu/2322/}

\bibitem{Hetal18}
J.-H. Huang, K.~Haffey, A.~Arefin, L.~E. Akhadov, J.~F. Harris, R.~Iyer,
  P.~Nath, {A microfluidic method to measure bulging heights for bulge testing
  of polydimethylsiloxane (PDMS) and polyurethane (PU) elastomeric membranes},
  RSC Adv. 8 (2018) 21133--21138.
\newblock \href {http://dx.doi.org/10.1039/C8RA01256C}
  {\path{doi:10.1039/C8RA01256C}}.

\bibitem{ZYSLYL08}
W.~Zhou, J.~Yang, G.~Sun, X.~Liu, F.~Yang, J.~Li, {Fracture properties of
  silicon carbide thin films by bulge test of long rectangular membrane}, J.
  Microelectromechan. Syst. 17 (2008) 453--461.
\newblock \href {http://dx.doi.org/10.1109/JMEMS.2008.916332}
  {\path{doi:10.1109/JMEMS.2008.916332}}.

\bibitem{ZYLJYY09}
W.~Zhou, J.~Yang, Y.~Li, A.~Ji, F.~Yang, Y.~Yu, {Bulge testing and fracture
  properties of plasma-enhanced chemical vapor deposited silicon nitride thin
  films}, Thin Solid Films 517 (2009) 1989--1994.
\newblock \href {http://dx.doi.org/10.1016/j.tsf.2008.10.042}
  {\path{doi:10.1016/j.tsf.2008.10.042}}.

\bibitem{YP02}
J.~Yang, O.~Paul, {Fracture properties of LPCVD silicon nitride thin films from
  the load-deflection of long membranes}, Sensors Actuators A: Physical 97-98
  (2002) 520--526.
\newblock \href {http://dx.doi.org/10.1016/S0924-4247(02)00049-3}
  {\path{doi:10.1016/S0924-4247(02)00049-3}}.

\bibitem{NHG12}
J.~Neggers, J.~P.~M. Hoefnagels, M.~G.~D. Geers, {On the validity regime of the
  bulge equations}, J. Mat. Res. 27 (2012) 1245--1250.
\newblock \href {http://dx.doi.org/10.1557/jmr.2012.69}
  {\path{doi:10.1557/jmr.2012.69}}.

\bibitem{SVOH18}
S.~Shafqat, O.~van~der Sluis, M.~Geers, J.~Hoefnagels, {A bulge test based
  methodology for characterizing ultra-thin buckled membranes}, Thin Solid
  Films 660 (2018) 88--100.
\newblock \href {http://dx.doi.org/10.1016/j.tsf.2018.04.005}
  {\path{doi:10.1016/j.tsf.2018.04.005}}.

\bibitem{YLMW14}
L.~Yang, S.-G. Long, Z.-S. Ma, Z.-H. Wang, {Accuracy analysis of plane-strain
  bulge test for determining mechanical properties of thin films}, Trans.
  Nonferrous Metals Soc. China 24 (2014) 3265--3273.
\newblock \href {http://dx.doi.org/10.1016/S1003-6326(14)63466-X}
  {\path{doi:10.1016/S1003-6326(14)63466-X}}.

\bibitem{ZPMSB98}
V.~Ziebart, O.~Paul, U.~M{\"{u}}nch, J.~Schwizer, H.~Baltes, {Mechanical
  properties of thin films from the load deflection of long clamped plates}, J.
  Microelectromechan. Syst. 7 (1998) 320--327.
\newblock \href {http://dx.doi.org/10.1109/84.709651}
  {\path{doi:10.1109/84.709651}}.

\bibitem{DHTJ17}
L.~Duclou{\'{e}}, A.~L. Hazel, A.~B. Thompson, A.~Juel, {Reopening modes of a
  collapsed elasto-rigid channel}, J. Fluid Mech. 819 (2017) 121--146.
\newblock \href {http://dx.doi.org/10.1017/jfm.2017.162}
  {\path{doi:10.1017/jfm.2017.162}}.

\bibitem{BGB18}
E.~Boyko, R.~Eshel, K.~Gommed, A.~D. Gat, M.~Bercovici, {Elastohydrodynamics of
  a pre-stretched finite elastic sheet lubricated by a thin viscous film with
  application to microfluidic soft actuators}, J. Fluid Mech. 862 (2019)
  732--752.
\newblock \href {http://dx.doi.org/10.1017/jfm.2018.967}
  {\path{doi:10.1017/jfm.2018.967}}.

\bibitem{DS16}
C.~Duprat, H.~A. Stone (Eds.), {Fluid--Structure Interactions in
  Low-Reynolds-Number Flows}, The Royal Society of Chemistry, Cambridge, UK,
  2016.
\newblock \href {http://dx.doi.org/10.1039/9781782628491}
  {\path{doi:10.1039/9781782628491}}.

\bibitem{GEGJ06}
T.~Gervais, J.~El-Ali, A.~G{\"{u}}nther, K.~F. Jensen, {Flow-induced
  deformation of shallow microfluidic channels}, Lab Chip 6 (2006) 500--507.
\newblock \href {http://dx.doi.org/10.1039/b513524a}
  {\path{doi:10.1039/b513524a}}.

\bibitem{CCSS17}
I.~C. Christov, V.~Cognet, T.~C. Shidhore, H.~A. Stone, {Flow rate--pressure
  drop relation for deformable shallow microfluidic channels}, J. Fluid Mech.
  814 (2018) 267--286.
\newblock \href {http://dx.doi.org/10.1017/jfm.2018.30}
  {\path{doi:10.1017/jfm.2018.30}}.

\bibitem{MY19}
A.~Mehboudi, J.~Yeom, {Experimental and theoretical investigation of a
  low-Reynolds-number flow through deformable shallow microchannels with
  ultra-low height-to-width aspect ratios}, Microfluid. Nanofluid. 23 (2019)
  66.
\newblock \href {http://dx.doi.org/10.1007/s10404-019-2235-9}
  {\path{doi:10.1007/s10404-019-2235-9}}.

\bibitem{OYE13}
O.~Ozsun, V.~Yakhot, K.~L. Ekinci, {Non-invasive measurement of the pressure
  distribution in a deformable micro-channel}, J. Fluid Mech. 734 (2013) R1.
\newblock \href {http://dx.doi.org/10.1017/jfm.2013.474}
  {\path{doi:10.1017/jfm.2013.474}}.

\bibitem{R45}
E.~Reissner, {The effect of transverse shear deformation on the bending of
  elastic plates}, ASME J. Appl. Mech. 12 (1945) A68--A77.

\bibitem{M51}
R.~D. Mindlin, {Influence of rotatory inertia and shear on flexural motions of
  isotropic, elastic plates}, ASME J. Appl. Mech. 18 (1951) 31--38.

\bibitem{CE19}
N.~Challamel, I.~Elishakoff, {A brief history of first-order shear-deformable
  beam and plate models}, Mech. Res. Commun. 102 (2019) 103389.
\newblock \href {http://dx.doi.org/10.1016/j.mechrescom.2019.06.005}
  {\path{doi:10.1016/j.mechrescom.2019.06.005}}.

\bibitem{ADC18}
V.~Anand, J.~David~JR, I.~C. Christov, {Non-Newtonian fluid--structure
  interactions: Static response of a microchannel due to internal flow of a
  power-law fluid}, J. Non-Newtonian Fluid Mech. 264 (2019) 62--72.
\newblock \href {http://dx.doi.org/10.1016/j.jnnfm.2018.12.008}
  {\path{doi:10.1016/j.jnnfm.2018.12.008}}.

\bibitem{HKO09}
P.~Howell, G.~Kozyreff, J.~Ockendon, {Applied Solid Mechanics}, Cambridge
  University Press, Cambridge, UK, 2009.
\newblock \href {http://dx.doi.org/10.1017/CBO9780511611605}
  {\path{doi:10.1017/CBO9780511611605}}.

\bibitem{reddy04}
J.~N. Reddy, {An Introduction to Nonlinear Finite Element Analysis}, Oxford
  University Press, Oxford, UK, 2004.

\bibitem{T21}
S.~P. Timoshenko, {On the correction for shear of the differential equation for
  transverse vibrations of prismatic bars}, Phil. Mag., Ser. 6 41~(245) (1921)
  744--746.
\newblock \href {http://dx.doi.org/10.1080/14786442108636264}
  {\path{doi:10.1080/14786442108636264}}.

\bibitem{GW01}
F.~Gruttmann, W.~Wagner, {Shear correction factors in Timoshenko's beam theory
  for arbitrary shaped cross-sections}, Comput. Mech. 27 (2001) 199--207.
\newblock \href {http://dx.doi.org/10.1007/s004660100239}
  {\path{doi:10.1007/s004660100239}}.

\bibitem{H01}
J.~R. Hutchinson, {Shear coefficients for Timoshenko beam theory}, ASME J.
  Appl. Mech. 68 (2001) 87--92.
\newblock \href {http://dx.doi.org/10.1115/1.1349417}
  {\path{doi:10.1115/1.1349417}}.

\bibitem{Z06}
S.~Zhang, {On the accuracy of Reissner--Mindlin plate model for stress boundary
  conditions}, ESAIM: M2AN 40 (2006) 269--294.
\newblock \href {http://dx.doi.org/10.1051/m2an:2006014}
  {\path{doi:10.1051/m2an:2006014}}.

\bibitem{SC18}
T.~C. Shidhore, I.~C. Christov, {Static response of deformable microchannels: a
  comparative modelling study}, J. Phys.: Condens. Matter 30 (2018) 054002.
\newblock \href {http://dx.doi.org/10.1088/1361-648X/aaa226}
  {\path{doi:10.1088/1361-648X/aaa226}}.

\bibitem{B10}
J.~Blaauwendraad, {Plates and FEM: Surprises and Pitfalls}, Vol. 171 of Solid
  Mechanics and Its Applications, Springer, Dordrecht, 2010.
\newblock \href {http://dx.doi.org/10.1007/978-90-481-3596-7}
  {\path{doi:10.1007/978-90-481-3596-7}}.

\bibitem{Love44}
A.~E.~H. Love, {A treatise on mathematical theory of elasticity}, 4th Edition,
  Dover Publications, New York, 1944.

\bibitem{TWK59}
S.~Timoshenko, S.~Woinowsky-Krieger, {Theory of Plates and Shells}, 2nd
  Edition, McGraw-Hill, New York, 1959.

\bibitem{reddy07}
J.~N. Reddy, {Theory and Analysis of Elastic Plates and Shells}, 2nd Edition,
  CRC Press, an imprint of Taylor {\&} Francis Group, Boca Raton, FL, 2007.

\bibitem{L07}
L.~G. Leal, {Advanced Transport Phenomena: Fluid Mechanics and Convective
  Transport Processes}, Cambridge University Press, New York, NY, 2007.
\newblock \href {http://dx.doi.org/10.2277/0521849101}
  {\path{doi:10.2277/0521849101}}.

\bibitem{ANSYS3}
{ANSYS Inc.}, {ANSYS{\textregistered} Academic Research Mechanical, Release
  19R2 Help System, Coupled Field Analysis Guide, ANSYS, Inc.}, Tech. rep.
  (2019).

\bibitem{CPFY12}
D.~Chakraborty, J.~R. Prakash, J.~Friend, L.~Yeo, {Fluid-structure interaction
  in deformable microchannels}, Phys. Fluids 24 (2012) 102002.
\newblock \href {http://dx.doi.org/10.1063/1.4759493}
  {\path{doi:10.1063/1.4759493}}.

\bibitem{PB99}
O.~Paul, H.~Baltes, {Mechanical behavior and sound generation efficiency of
  prestressed, elastically clamped and thermomechanically driven thin film
  sandwiches}, J. Micromech. Microeng. 9 (1999) 19--29.
\newblock \href {http://dx.doi.org/10.1088/0960-1317/9/1/002}
  {\path{doi:10.1088/0960-1317/9/1/002}}.

\bibitem{CC15}
S.~C. Chapra, R.~P. Canale, {Numerical Methods for Engineers}, 7th Edition,
  McGraw-Hill Education, New York, NY, 2015.

\bibitem{SciPy}
P.~Virtanen, R.~Gommers, T.~E. Oliphant, M.~Haberland, T.~Reddy, D.~Cournapeau,
  E.~Burovski, P.~Peterson, W.~Weckesser, J.~Bright, S.~J. van~der Walt,
  M.~Brett, J.~Wilson, K.~J. Millman, N.~Mayorov, A.~R.~J. Nelson, E.~Jones,
  R.~Kern, E.~Larson, C.~Carey, {\.I}.~Polat, Y.~Feng, E.~W. Moore,
  J.~VanderPlas, D.~Laxalde, J.~Perktold, R.~Cimrman, I.~Henriksen, E.~A.
  Quintero, C.~R. Harris, A.~M. Archibald, A.~H. Ribeiro, F.~Pedregosa, P.~van
  Mulbregt, {SciPy 1. 0 Contributors},
  \href{https://arxiv.org/abs/1907.10121}{{SciPy 1.0--Fundamental Algorithms
  for Scientific Computing in Python}}, preprint.
\newline\urlprefix\url{https://arxiv.org/abs/1907.10121}

\bibitem{M84}
J.~Mandel, {The Statistical Analysis of Experimental Data}, Dover Publications,
  Mineola, NY, 1984.

\bibitem{RS16}
A.~Raj, A.~K. Sen, {Flow-induced deformation of compliant microchannels and its
  effect on pressure--flow characteristics}, Microfluid. Nanofluid. 20 (2016)
  31.
\newblock \href {http://dx.doi.org/10.1007/s10404-016-1702-9}
  {\path{doi:10.1007/s10404-016-1702-9}}.

\bibitem{KKJ11}
T.~K. Kim, J.~K. Kim, O.~C. Jeong, {Measurement of nonlinear mechanical
  properties of PDMS elastomer}, Microelectron. Eng. 88 (2011) 1982--1985.
\newblock \href {http://dx.doi.org/10.1016/j.mee.2010.12.108}
  {\path{doi:10.1016/j.mee.2010.12.108}}.

\bibitem{CVB08}
J.~Chopin, D.~Vella, A.~Boudaoud, {The liquid blister test}, Proc. R. Soc. A
  464 (2008) 2887--2906.
\newblock \href {http://dx.doi.org/10.1098/rspa.2008.0095}
  {\path{doi:10.1098/rspa.2008.0095}}.

\bibitem{MCSPS19}
A.~Mart{\'{i}}nez-Calvo, A.~Sevilla, G.~G. Peng, H.~A. Stone, {Start-up flow in
  shallow deformable microchannels}, J. Fluid Mech. 885 (2020) A25.
\newblock \href {http://dx.doi.org/10.1017/jfm.2019.994}
  {\path{doi:10.1017/jfm.2019.994}}.

\bibitem{HBDB14}
I.~J. Hewitt, N.~J. Balmforth, J.~R. De~Bruyn, {Elastic-plated gravity
  currents}, Eur. J. Appl. Math. 26 (2015) 1--31.
\newblock \href {http://dx.doi.org/10.1017/S0956792514000291}
  {\path{doi:10.1017/S0956792514000291}}.

\bibitem{MHT15}
F.~Meng, J.~Huang, M.~D. Thouless, {The collapse and expansion of liquid-filled
  elastic channels and cracks}, ASME J. Appl. Mech. 82 (2015) 101009.
\newblock \href {http://dx.doi.org/10.1115/1.4031048}
  {\path{doi:10.1115/1.4031048}}.

\bibitem{EG16}
S.~B. Elbaz, A.~D. Gat, {Axial creeping flow in the gap between a rigid
  cylinder and a concentric elastic tube}, J. Fluid Mech. 806 (2016) 580--602.
\newblock \href {http://dx.doi.org/10.1017/jfm.2016.587}
  {\path{doi:10.1017/jfm.2016.587}}.

\bibitem{BN18}
T.~V. Ball, J.~A. Neufeld, {Static and dynamic fluid-driven fracturing of
  adhered elastica}, Phys. Rev. Fluids 3 (2018) 074101.
\newblock \href {http://dx.doi.org/10.1103/PhysRevFluids.3.074101}
  {\path{doi:10.1103/PhysRevFluids.3.074101}}.

\end{thebibliography}


\clearpage
\appendix
\setcounter{equation}{0}
\setcounter{page}{1}
\lfoot{\textsf{Supplemental Material, JAM-19-1506}}
\rfoot{\textsf{Page \thepage}}
\renewcommand\theequation{A\arabic{equation}}

\centerline{\Large\textbf{Supplemental Material: JAM-19-1506}}

\medskip

\centerline{\large V.~Anand, S.~C.~Muchandimath, I.~C.~Christov}

\balance

\section{Plate Theory Notation, Definitions, and Derivations}
\label{app:plate_theory}

\subsection{Kinematics}

In FOSDT, the assumption that the transverse normals are straight and inextensible leads to the following displacement field within the plate \cite{reddy07}: 
\begin{subequations}\label{eq:displacements_x_z}\begin{align}
    u_x (x,y,z) &= u_{x0}(x,z) + y\phi_x(x,z), \\
    u_z(x,y,z) & = u_{z0}(x,z) + y\phi_z(x,z), \\
    \label{eq:displacement_y} u_y(x,z,y) &= u_{y0}(x,z).
\end{align}\end{subequations}
Here, $u_{x0}$ and $u_{z0}$ are the in-plane displacements, $u_{y0}$ is the transverse displacement (henceforth denoted just as $u_{y}$ for simplicity and without fear of confusion), and $\phi_x$ and $\phi_z$ are the rotations of the normal to the plate about the $x$- and the $z$-axis, respectively. Equations~\eqref{eq:displacements_x_z} are written assuming $y =0$ is the mid-plane (neutral surface) of the plate, as shown in Fig.~\ref{fig:schematic}. In FOSDT, the nonlinear terms in the strain tensor, which arise from von K\'arm\'an strains, are neglected and the strain tensor is written in the column vector form as:
\begin{equation}
\label{eq:strain_tensor_linear}
     \underbrace{ \begin{pmatrix}
    \gamma_{xx} \\
    \gamma_{zz} \\
    \gamma_{yy} \\
    \gamma_{xz} \\
    \gamma_{xy} \\
    \gamma_{zy} \\
    \end{pmatrix}}_{\doubleunderline{\gamma}}
    = \underbrace{\begin{pmatrix}
    \frac{\partial u_{x0}}{\partial x} \\[1mm]
    \frac{\partial u_{z0}}{\partial z} \\
    0 \\
    \frac{\partial u_{x0}}{\partial z}+\frac{\partial u_{z0}}{\partial x} \\[1mm]
    \frac{\partial u_{y}}{\partial x}+\phi_x \\[1mm]
    \frac{\partial u_{y}}{\partial z}+\phi_z \\
    \end{pmatrix}}_{\doubleunderline{\gamma}^0}
    \; +\; y \; \underbrace{\begin{pmatrix}
    \frac{\partial \phi_x}{\partial x} \\[1mm]
    \frac{\partial \phi_z}{\partial z} \\
    0 \\
    \frac{\partial \phi_x}{\partial z}+ \frac{\partial \phi_z}{\partial x} \\
    0\\
    0
    \end{pmatrix}}_{\doubleunderline{\varPhi}}
\end{equation}
Here, $\doubleunderline{\varPhi}$ is the curvature strain tensor, which arises from bending, while $\doubleunderline{\gamma}^0$ represents the in-plane stretching and deformation due to transverse shear. It is a feature of FOSDT that the transverse shear strains remain constant across the thickness, while the in-plane strains vary linearly with $y$.

\subsection{Equations of Static Equilibrium}
As is standard in plate theory, we integrate the stresses across the thickness and define the corresponding stress resultants as
\begin{equation}
\label{eq:stress_resultants_normal_defined}
    \begin{pmatrix}
    N_{xx} \\
    N_{zz}\\
    N_{xz}\\
    Q_{x}\\
    Q_{z}\\
\end{pmatrix}    
=\int_{-t/2}^{+t/2}\begin{pmatrix}
\sigma_{xx} \\
\sigma_{zz}\\
\sigma_{xz} \\
\sigma_{xy} \\
\sigma_{zy}
\end{pmatrix} \mathrm{d}y,
\end{equation}
and the bending moments as
\begin{equation}
\label{eq:bending_moment_defined}
    \begin{pmatrix}
    M_{xx}\\
    M_{zz}\\
    M_{xz}
    \end{pmatrix}
    =\int_{-t/2}^{+t/2}
    \begin{pmatrix}
    \sigma_{xx}\\
    \sigma_{zz}\\
    \sigma_{xz}
    \end{pmatrix} y \,\mathrm{d}y.
\end{equation}
There are only two independent variables in the plate theory: the in-plane coordinates $x$ and $z$. Thus, here, $N_{xx}$ and $N_{zz}$ are the normal stress resultants in these, $x$ and $z$ directions, respectively. Likewise, $Q_x$ and $Q_z$ are the transverse shear stress resultants acting on the planes which have their outward normals in the $x$ and $z$ directions, respectively. Meanwhile, $N_{xz}$ is the in-plane shear stress resultant. Similarly, $M_{xx}$ and $M_{zz}$ are bending moments, while $M_{xz}$ is the twisting moment. There are no moments due to the transverse stresses $\sigma_{xy}$ and $\sigma_{zy}$. Additionally, the assumption of a plane-stress state means that $\sigma_{yy} = 0$, and $\sigma_{yy}$ does not contribute to any stress resultants.

The equations of equilibrium, written in terms of the stress resultants \cite[Ch.~10]{reddy07}, are 
\begin{subequations}\begin{align}
\label{eq:equil_in_plane_1}
\frac{\partial N_{xx}}{\partial x} + \frac{\partial N_{xz}}{\partial z} &= 0, \displaybreak[3]\\
\label{eq:equil_in_plane_2}
\frac{\partial N_{xz}}{\partial x} + \frac{\partial N_{zz}}{\partial z} &= 0, \displaybreak[3]\\
\label{eq:equil_trans}
\frac{\partial Q_x}{\partial x} + \frac{\partial Q_z}{\partial z}+\mathcal{N} + p &= 0, \displaybreak[3]\\
\label{eq:equil_moment_1}
\frac{\partial M_{xx}}{\partial x} + \frac{\partial M_{xz}}{\partial z} - Q_x &= 0, \displaybreak[3]\\
\label{eq:equil_moment_2}
\frac{\partial M_{xz}}{\partial x} + \frac{\partial M_{zz}}{\partial z} - Q_z &= 0,
\end{align}\label{eq:equil}\end{subequations}
where
\begin{equation}
\label{eq:mathcal_N}
    \mathcal{N} := \frac{\partial}{\partial x}\left(N_{xx}\frac{\partial u_{y}}{\partial x}+N_{xz}\frac{\partial u_{y}}{\partial z}\right)+\frac{\partial }{\partial z}\left(N_{xz}\frac{\partial u_{y}}{\partial x}+N_{zz}\frac{\partial u_{y}}{\partial z}\right)
\end{equation}
couples the displacement in the transverse direction (bending) to the in-plane displacements (stretching). This term accounts for moderate rotations and originates from employing von K\'arm\'an strains in the derivation of the equations of equilibrium \cite{reddy07}. 

Thus, we have neglected the nonlinear terms in the kinematics of the problem, but opted to retain these terms in the equations of static equilibrium. Neglecting $\mathcal{N}$ in Eq.~\eqref{eq:equil_trans}, would decouple the bending response from the stretching response. In other words, the transverse deflection would not be affected by stretching (pre-stress) at all, which is valid only when the stretching is negligible. Retaining $\mathcal{N}$ in the equations of static equilibrium thus enlarges the scope of application of the theory, and allows for the consideration of pre-stressed (pre-stretched) plates. On the other hand, if we had also incorporated the nonlinear (moderate rotation) terms in the kinematics, and employed the von K\'arm\'an strains, then we would have obtained the \emph{nonlinear} von K\'arm\'an plate theory \cite{HKO09}, which is difficult (if not impossible) to solve analytically \cite{HKO09,reddy07}. In the von K\'arm\'an plate theory, stretching and bending responses are tightly coupled, unlike a linear plate theory in which $\mathcal{N}$ is dropped altogether from the analysis. 

To summarize: in this paper, the coupling between stretching and bending is one-way; stretching influences bending but the converse is not true. {The influence of stretching in the bending response is accounted for by incorporating $\mathcal{N}$, given by Eq.~\eqref{eq:mathcal_N}, which appears in Eq.~\eqref{eq:equil_trans}. On the other hand, however, Eqs.~\eqref{eq:equil_in_plane_1} and \eqref{eq:equil_in_plane_2} that govern the in-plane equilibrium (stretching) are decoupled from Eqs.~\eqref{eq:equil_trans}--\eqref{eq:equil_moment_2} and do not contain any terms corresponding to the bending response.} The current theory may thus be regarded as ``weakly nonlinear'' in a sense, providing a suitable trade-off between the nonlinear von K\'arm\'an plate theory (with stretching) and a linear FOSDT theory (in which stretching decouples).
 
\subsection{Constitutive Equations}
For the condition of plane stress, the constitutive equations reduce (see \cite{B10,reddy07}) to 
 \begin{equation}
 \label{eq:consti_stress}
     \begin{pmatrix}
     \sigma_{xx} \\
     \sigma_{zz} \\
     \sigma_{xz}
     \end{pmatrix}
     =\frac{E}{(1-\nu^2)}\begin{pmatrix}
     1& \nu & 0 \\
     \nu & 1 & 0 \\
     0 & 0 & \frac{1-\nu}{2}
     \end{pmatrix}
     \begin{pmatrix}
     \gamma_{xx} \\
     \gamma_{zz} \\
     \gamma_{xz}
     \end{pmatrix},
 \end{equation}
where $\nu$ is the Poisson ratio and $E$ is the Young's modulus of the linearly elastic material. Next, we substitute the expressions for the strains in terms of displacements from Eq.~\eqref{eq:strain_tensor_linear} into Eq.~\eqref{eq:consti_stress}, the result of which, upon being employed in Eq.~\eqref{eq:stress_resultants_normal_defined}, yields:
\begin{equation}
\label{eq:normal_stress_consti}
 \begin{pmatrix}
    N_{xx} \\
    N_{zz}\\
    N_{xz}\\
\end{pmatrix}
= \frac{Et}{(1-\nu^2)}\begin{pmatrix}
     1& \nu & 0 \\
     \nu & 1 & 0 \\
     0 & 0 & \frac{1-\nu}{2}
     \end{pmatrix}
     \begin{pmatrix}
    \frac{\partial u_{x0}}{\partial x} \\[1mm]
    \frac{\partial u_{z0}}{\partial z} \\[1mm]
    \frac{\partial u_{x0}}{\partial z}+\frac{\partial u_{z0}}{\partial x} \\
     \end{pmatrix}.
\end{equation}
Note that due to the assumption about linear strains, the in-plane stress resultants are only functions of the in-plane strains, and they are independent of the transverse deflections and rotations. Similarly, the bending moments from Eq.~\eqref{eq:bending_moment_defined} are calculated to be
\begin{equation}
\label{eq:bending_moment_constitutive2}
    \begin{pmatrix}
    M_{xx}\\
    M_{zz}\\
    M_{xz}
\end{pmatrix}= \frac{Et^3}{12(1-\nu^2)}\begin{pmatrix}
     1& \nu & 0 \\
     \nu & 1 & 0 \\
     0 & 0 & \frac{1-\nu}{2}
     \end{pmatrix} \begin{pmatrix}
    \frac{\partial \phi_x}{\partial x} \\[1mm]
    \frac{\partial \phi_z}{\partial z} \\[1mm]
    \frac{\partial \phi_x}{\partial z}+ \frac{\partial \phi_z}{\partial x} \\
     \end{pmatrix},
\end{equation}
where we observe that the bending moments are only a function of the rotations.

Next, the constitutive equations for the shear stresses are modified as:
\begin{subequations}\begin{align}
    \sigma_{xy} &\approx \kappa G\gamma_{xy} \\
    \sigma_{zy} &\approx \kappa G\gamma_{zy},
\end{align}\label{eq:consti_shear_stress}\end{subequations}
where $G = E/[2(1+\nu)]$ is the shear modulus, and  $\kappa$ is Timoshenko's ``shear correction factor'' \cite{T21}, which is  commonly introduced to account for nonuniform distribution of the transverse shear strain across the thickness \cite{GW01,H01,Z06}. Now, we substitute the expressions for $\sigma_{xy}$ and $\sigma_{zy}$ from Eqs.~\eqref{eq:consti_shear_stress} into Eq.~\eqref{eq:stress_resultants_normal_defined} to relate the shear stress resultants to the deformation and rotation of the normal:
\begin{subequations}\label{eq:transverse_shear_res_kappa}\begin{align}
Q_x &= \kappa\int_{-t/2}^{+t/2}G\gamma_{xy}\,\mathrm{d}y = \kappa Gt\left[\frac{\partial u_{y}}{\partial x}+\phi_x\right],\\
Q_z &= \kappa\int_{-t/2}^{+t/2}G\gamma_{zy}\,\mathrm{d}y = \kappa Gt\left[\frac{\partial u_{y}}{\partial z}+\phi_z\right].
\end{align}\end{subequations}

Zhang \cite{Z06} proved mathematically  that the equations of linear elasticity and those of the RM plate theory, both in the limit of $t/w \to 0$, agree only when $\kappa = 1$. Therefore, as in our previous works \cite{SC18,ADC18}, we take $\kappa = 1$ when generating our results below. However, we keep the variable $\kappa$ throughout our equations for consistency with the applied mechanics literature. This completes the derivation of the stress resultants in terms of the displacements under the FOSDT.

\balance

\clearpage
\section{Deformation Profile in Other Regimes}
\label{app:other_regimes}

\subsection{Regime 1}
\label{app:Regime1}

In this regime, as $\lambda \ll 1$, we simply take the formal limit $\lambda \to 0$, and Eq.~\eqref{eq:dimless_equilibrium_z_v8} reduces to
\begin{equation}
\label{eq:dimless_equilibrium_momentum_regime_1}
    \frac{1}{\mathscr{T}} \left(\frac{\partial \Phi_X}{\partial X}+\frac{\partial ^2 U}{\partial X^2 }\right) = - \frac{w^4\mathcal{P}_c}{D_b \mathcal{U}_c}P(Z).
\end{equation}
Eliminating $\Phi_X$ between Eqs.~\eqref{eq:dimless_equilibrium_momentum_regime_1} and \eqref{eq:dimless_equilibrium_z_v7}, we obtain a single ODE in $U$:
\begin{equation}
    \label{eq:ODE_regime_1_1}
    \frac{\partial^4 U}{\partial X^4} = \frac{w^4\mathcal{P}_c}{D_b \mathcal{U}_c}P(Z).
 \end{equation}
Clearly, the appropriate choice for the deformation scale is
\begin{equation}
    \mathcal{U}_c =\frac{w^4\mathcal{P}_c}{D_b},
\end{equation}
and, therefore, Eq.~\eqref{eq:ODE_regime_1_1} becomes:
\begin{equation}
     \label{eq:ODE_regime_1_2}
    \frac{\partial^4 U}{\partial X^4} = P(Z).
\end{equation} 

Next, the BCs $\Phi_X|_{X= \pm 1/2}=0$ need to be converted to appropriate BCs on $U$. To this end, differentiate Eq.~\eqref{eq:dimless_equilibrium_momentum_regime_1} with respect to $X$ to obtain an expression for ${\partial ^2 \Phi_x}/{\partial X^2}$, which can be evaluated at $X=\pm 1/2$. Then, substituting the latter, along with imposing  $\Phi_X|_{X=\pm1/2} =0$, into Eq.~\eqref{eq:dimless_equilibrium_z_v7} evaluated at $X=\pm 1/2$ yields:
\begin{equation}
\label{eq:boundary_conditions_regime_1}
\left.\left(\frac{\partial U}{\partial X} + \frac{1}{\mathscr{T}}\frac{\partial^3 U}{\partial X^3}\right)\right|_{X=\pm1/2} = 0.
\end{equation}

The solution of Eq.~\eqref{eq:ODE_regime_1_2} subject to Eq.~\eqref{eq:boundary_conditions_regime_1} and $U|_{X=\pm1/2}=0$ is easily found to be
\begin{equation}
\label{eq:U_XZ_Tanmay}
   U(X,Z) = \frac{P(Z)}{24}\left[X^4 - \left(\mathscr{T}+\frac{1}{2}\right)X^2 + \frac{1}{4}\left( \mathscr{T} + \frac{1}{4}\right) \right].
\end{equation}
This profile was already obtained in \cite[Eq.~(21)]{SC18} in the absence of pre-tension.

\subsection{Regime 3a}
\label{app:Regime3a}

For Regime 3a, $\lambda = \mathcal{O}( 1/\delta^2)$, thus $\lambda \gg 1$ for $\delta \ll 1$.  Keeping only the largest terms in Eq.~\eqref{eq:regime_2_ODE_v1} for $\lambda\gg1$, we obtain
\begin{equation}
    \label{eq:regime3_ode_v2}
    \mathscr{T}\frac{\partial^4 U}{\partial X^4} - \frac{\partial^2 U}{\partial X^2} = \frac{1}{\lambda \delta^2} P(Z).
\end{equation}
Retaining the pressure on the right-hand side of the last equation can be justified by arguing that, in Regime 3a, we should choose the scale deformation
\begin{equation}
\label{eq:regime3_u_scale}
    \mathcal{U}_c = \frac{h_0^2w^2\mathcal{P}_c}{D_b},
\end{equation}
so that the coefficient of $P(Z)$ is now $1/(\lambda\delta^2)=\mathcal{O}(1)$.

The ODE~\eqref{eq:regime3_ode_v2} is still subject to the boundary conditions given in Eq.~\eqref{eq:clamping_BCs}, which means that, as before, we need to convert the BCs on $\Phi_X$ to appropriate BCs on $U$. To that end, we first insert $\Phi_X|_{X=\pm1/2} = 0$ in Eq.~\eqref{eq:dimless_equilibrium_z_v7} to obtain:
\begin{equation}
\label{eq:regime3_phi_bc1}
    \left.\left(\mathscr{T}\frac{\partial ^2 \Phi_X}{\partial X^2}-\frac{\partial U}{\partial X} \right)\right|_{X=\pm1/2} = 0.
\end{equation}
Next, we differentiate Eq.~\eqref{eq:dimless_equilibrium_z_v8}, evaluate it at $X=\pm1/2$, and insert the expression for ${\partial^2 \Phi_X}/{\partial X^2}$ from Eq.~\eqref{eq:regime3_phi_bc1} into it to obtain:
\begin{equation}
    \left.\left[\left(\frac{\delta^2}{\mathscr{T}}\right)\frac{\partial U}{\partial X}+(\lambda\mathscr{T}+1)\delta^2\frac{\partial ^3 U}{\partial X^3}\right]\right|_{X=\pm1/2}  = 0.
\end{equation}
Now, since $\lambda = \mathcal{O}(1/\delta^2)$, the above equation to the leading order in $\delta\ll1$ is
\begin{equation}
\label{eq:regime3_mixed_bc}
    \left.\frac{\partial^3U}{\partial X^3}\right|_{X=\pm1/2}  = 0.
\end{equation}

The solution of Eq.~\eqref{eq:regime3_ode_v2} subject to Eq.~\eqref{eq:regime3_mixed_bc} and $U|_{X=\pm1/2}=0$ is easily found to be
\begin{equation}
\label{eq:regime3_final}
U(X,Z) =\frac{P(Z)}{2\lambda\delta^2}\left(\frac{1}{4}-{X^2}\right),
\end{equation}
where, due to the differing choice in $\mathcal{U}_c$, the last equation contains a $\delta^2$ not present in Eq.~\eqref{eq:xsect_sol_FSDT_strech_regime3_v4}. Note that neglecting bending rendered the deformation profile given in Eq.~\eqref{eq:regime3_final} independent of the thickness of the structure. Thus, Eq.~\eqref{eq:regime3_final} is suitable for both thin and thick plates. Equation~\eqref{eq:regime3_final} has been used in the literature to characterize the material properties of thin membranes undergoing strong compression ($\lambda < 0$) and buckling (see Eq.~(8) and Table II in \cite{ZPMSB98}).

\clearpage
\section{Thin-Plate Theory}
\label{app:thin}

For a Kirchhoff--Love \cite{Love44,TWK59} or ``classical'' \cite{reddy07} (thin) plate theory, $\phi_x$ and $\phi_z$ are not independent degrees of freedom; instead, they are expressed in terms of the transverse displacement (see, e.g., \cite[Ch.~4]{B10}) as
\begin{equation}
\label{eq:rotation_thin_plate}
    \phi_x =- \frac{\partial u_{y}}{\partial x}, \qquad
    \phi_z =- \frac{\partial u_{y}}{\partial z}.
\end{equation}
The constitutive equations for the in-plane stress resultants, $N_{xx},N_{zz}$ and $N_{xz}$ for the thin-plate theory are still given by Eq.~\eqref{eq:normal_stress_consti}. However, for the bending moments, when we substitute the equations for the rotations from Eqs.~\eqref{eq:rotation_thin_plate} into Eq.~\eqref{eq:bending_moment_constitutive2}, we obtain:
\begin{equation}
\label{eq:bending_moment_constitutive_thin}
    \begin{pmatrix}
    M_{xx}\\
    M_{zz}\\
    M_{xz}
\end{pmatrix}
= \frac{-Et^3}{12(1-\nu^2)}
    \begin{pmatrix}
     1& \nu & 0 \\
     \nu & 1 & 0 \\
     0 & 0 & (1-\nu)/2
     \end{pmatrix} 
     \begin{pmatrix}
    \frac{\partial^2 u_{y}}{\partial x^2} \\[1mm]
    \frac{\partial^2 u_{y}}{\partial z^2} \\[1mm]
    2\frac{\partial^2 u_{y}}{\partial z \partial x} \\
     \end{pmatrix}.
\end{equation}

The equations for static equilibrium for a thin plate undergoing combined bending and stretching are the same as those for thick plate. However, for the thin-plate theory, since the transverse shear stains (and, thus, the transverse shear stress resultants) are negligible, it is customary to eliminate $Q_x$ and $Q_z$ from Eqs.~\eqref{eq:equil_trans}, \eqref{eq:equil_moment_1} and \eqref{eq:equil_moment_2} to obtain:
\begin{equation}
    \label{eq:equi_thin}
    \frac{\partial ^2 M_{xx}}{\partial x^2}+\frac{\partial ^2 M_{zz}}{\partial z^2}+2\frac{\partial ^2 M_{xz}}{\partial x\partial z}+\mathcal{N}=0.
\end{equation}
Equation \eqref{eq:equi_thin} along with Eqs.~\eqref{eq:equil_in_plane_1} and \eqref{eq:equil_in_plane_2} are the equations expressing the static equilibrium (see also \cite[Ch.~3]{reddy07}).
   
Equations~\eqref{eq:displacement_in_plane_x} and \eqref{eq:displacement_in_plane_z} are still the equations governing the in-plane displacements for the thin-plate case. To obtain a PDE governing the transverse deflection of the thin plate, we substitute the bending moments from Eq.~\eqref{eq:bending_moment_constitutive_thin} into Eq.~\eqref{eq:equi_thin} to obtain:
\begin{multline}
    \label{eq:thin_plate_stretching_deformation}
    N_{xx}\frac{\partial^2u_{y}}{\partial x^2}+2N_{xz}\frac{\partial^2 u_{y}}{\partial x\partial z}+N_{zz}\frac{\partial ^2 u_{y}}{\partial z^2}\\
    -\frac{Et^3}{12(1-\nu^2)}\left(\frac{\partial^4u_{y}}{\partial x^4}+\frac{\partial^4u_{y}}{\partial z^4}+2\frac{\partial^4u_{y}}{\partial x^2 \partial z^2}\right) + p = 0,
\end{multline}
which is  the ``plate equation'' in \cite[Eq.~(4.6.12)]{HKO09}. For a constant pre-tension as in Eq.~\eqref{eq:pre-tension-defined},  Eq.~\eqref{eq:thin_plate_stretching_deformation} becomes
\begin{equation}
    \label{eq:thin_plate_pre-tension}
    T\left(\frac{\partial^2u_{y}}{\partial x^2}+\frac{\partial ^2 u_{y}}{\partial z^2}\right) - D_b\left(\frac{\partial^4u_{y}}{\partial x^4} + \frac{\partial^4u_{y}}{\partial z^4} + 2\frac{\partial^4u_{y}}{\partial x^2 \partial z^2}\right) + p = 0,
\end{equation}
which is the same as \cite[Eq.~(4.6.14)]{HKO09}

Using the dimensionless variables from Eq.~\eqref{eq:scales},  Eq.~\eqref{eq:thin_plate_pre-tension} can be rewritten as
\begin{multline}
\label{eq:thin_plate_pre-tension_dimless}
    \lambda\left(\delta^4\frac{\partial ^2 U}{\partial X^2}+\epsilon^2\delta^2\frac{\partial ^2 U}{\partial Z^2}\right) \\
    -\left(\delta^4\frac{\partial ^4 U}{\partial X^4}+\epsilon^4\frac{\partial ^4 U}{\partial Z^4} +2\epsilon^2\delta^2\frac{\partial ^4 U}{\partial X^2{\partial Z^2}}\right)
    +\frac{w^4\mathcal{P}_c}{\mathcal{U}_c D_b}\delta^4 P = 0.
\end{multline}
To the leading order in $\epsilon$, Eq.~\eqref{eq:thin_plate_pre-tension_dimless} is
\begin{equation}
\label{eq:thin_plate_pre-tension_dimless2}
    \lambda\frac{\partial ^2 U}{\partial X^2} - \frac{\partial ^4 U}{\partial X^4} + \frac{w^4\mathcal{P}_c}{\mathcal{U}_cD_b} P = 0.
\end{equation}
Choosing the scale of deformation $\mathcal{U}_c$ as in  Eq.~\eqref{eq:deformation_scale} balances all terms and reduces Eq.~\eqref{eq:thin_plate_pre-tension_dimless2} to:
\begin{equation}
\label{eq:thin_plate_pre-tension_dimless3}
    \lambda\frac{\partial ^2 U}{\partial X^2}-\frac{\partial ^4 U}{\partial X^4} + {P}(Z) = 0.
\end{equation}
Obviously, Eq.~\eqref{eq:thin_plate_pre-tension_dimless3} can also be obtained by taking $\mathscr{T} \to 0$ in Eq.~\eqref{eq:regime_2_ODE_v1}, which was derived from FOSDT.

Solving Eq.~\eqref{eq:thin_plate_pre-tension_dimless3} subject to thin-plate clamped BCs, i.e.,  $U|_{X=\pm 1/2} = (\partial U/\partial X)|_{X=\pm 1/2} = 0$, we obtain 
\begin{multline}
\label{eq:xsect_sol_thin_plate_strech}
    U(X,Z) = \\ \frac{P(Z)}{2\lambda} \left\{ \left(\frac{1}{4}- X^2\right) - \left[\frac{\cosh \left(\frac{1}{2}\sqrt{\lambda}\right) - \cosh \left(X\sqrt{\lambda}\right)}{\sqrt{\lambda} \sinh\left(\frac{1}{2}\sqrt{\lambda}\right)}\right]\right\}.
\end{multline}

\begin{figure}[b]
    \centering
    \includegraphics[width=\linewidth]{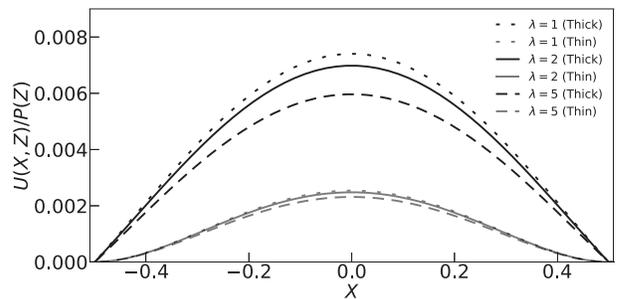}
    \caption{Ratio of the dimensionless deformation profile $U(X,Z)$ to the hydrodynamic pressure $P(Z)$ across the cross-section $X\in[-1/2,+1/2]$, for different values of the tension-to-bending parameter $\lambda$, as predicted by the thick-plate theory (FOSDT) from Eq.~\eqref{eq:xsect_sol_FSDT_strech} (black curves) and the thin-plate theory from Eq.~\eqref{eq:xsect_sol_thin_plate_strech} (grey curves).}
    \label{fig:ThinThick_Deformation}
\end{figure}

The deformation profile given by Eq.~\eqref{eq:xsect_sol_thin_plate_strech} has been previously derived in the literature to describe the deformations of a thin plate undergoing both bending and stretching \cite[Eq.~(8)]{ZPMSB98} (see also \cite[Eq.~(10)]{ZYLJYY09}), An example comparison between Eq.~\eqref{eq:xsect_sol_thin_plate_strech} and  Eq.~\eqref{eq:xsect_sol_FSDT_strech} is shown in Fig.~\ref{fig:ThinThick_Deformation}. As should be expected, the thick-plate theory predicts a larger deformation than the thin-plate theory, for the same pressure load. This observation is attributed to the fact that a thick plate can support deformation due to transverse shear, while a thin plate cannot.

The limit as $\lambda\to0$ is singular and must be taken carefully (e.g., with {\sc Mathematica}) to yield 
\begin{equation}
    U(X,Z) = \frac{P(Z)}{24}\left(\frac{1}{4} - X^2\right)^2,
\end{equation} 
which is the tension-free thin-plate result from  \cite{CCSS17}.

\end{document}